\documentclass[fleqn,usenatbib]{mnras}

\usepackage{newtxtext,newtxmath}

\usepackage[T1]{fontenc}

\DeclareRobustCommand{\VAN}[3]{#2}
\let\VANthebibliography\thebibliography
\def\thebibliography{\DeclareRobustCommand{\VAN}[3]{##3}\VANthebibliography}

\usepackage{graphicx}	
\usepackage{amsmath}	

\usepackage{amssymb}	

\usepackage{subcaption}

\newcommand{\meter}{\,\mathrm{m}}
\newcommand{\au}{\,\mathrm{au}}
\newcommand{\yr}{\,\mathrm{yr}}
\newcommand{\arcsecyr}{\,\mathrm{^{\prime\prime} yr^{-1}}}
\newcommand{\degree}{\mathrm{^\circ}}

\title[Resonant mechanisms producing near-Sun asteroids]{Resonant mechanisms that produce near-Sun asteroids}

\author[A. Toliou et al.]{Athanasia Toliou$^{1}$\thanks{Corresponding author; email: athanasia.toliou@ltu.se} and Mikael Granvik$^{1,2}$
\\
$^{1}$Asteroid Engineering Laboratory, Lule\r{a} University of Technology, Box 848, SE-98128 Kiruna, Sweden\\
$^{2}$Department of Physics, PO Box 64, 00014 University of Helsinki, Finland
}

\date{Accepted XXX. Received YYY; in original form ZZZ}

\pubyear{2022}

\begin{document}
\label{firstpage}
\pagerange{\pageref{firstpage}--\pageref{lastpage}}
\maketitle

\begin{abstract}
All near-Earth asteroids (NEAs) that reach sufficiently small perihelion distances will undergo a so-called super-catastrophic disruption. The mechanisms causing such disruptions are currently unknown or, at least, undetermined. To help guide theoretical and experimental work to understand the disruption mechanism, we use numerical simulations of a synthetic NEA population to identify the resonant mechanisms that are responsible for driving NEAs close to the Sun, determine how these different mechanisms relate to their dynamical lifetimes at small heliocentric distances and calculate the average time they spend at different heliocentric distances. Typically, resonances between NEAs and the terrestrial and giant planets are able to dramatically reduce the perihelion distances of the former. We developed an algorithm that scans the orbital evolution of asteroids and automatically identifies occurrences of mean motion and secular resonances. We find that most near-Sun asteroids are pushed to small perihelion distances by the 3:1J and 4:1J mean-motion resonances with Jupiter, as well as the secular resonances $\nu_6$, $\nu_5$, $\nu_3$ and $\nu_4$. The time-scale of the small-perihelion evolution is fastest for the 4:1J, followed by the 3:1J, while $\nu_5$ is the slowest. $\sim7$ per cent of the test asteroids were not trapped in a resonance during the latest stages of their dynamical evolution, which suggests that the secular oscillation of the eccentricity due to the Kozai mechanism, a planetary close encounter or a resonance that we have not identified pushed them below the estimated average disruption distance. 
\end{abstract}




\section{Introduction}

The main asteroid belt is considered to be the prime contributor of near-Earth asteroids (NEAs). Main-belt asteroids (MBAs) are continuously replenishing the NEA population through certain dynamical pathways. \citet{Granvik2017}, using dynamical simulations of MBAs, have studied virtually all of these possible escape regions (ER) in the asteroid belt. Based on those results, \citet{Granvik2018} developed a debiased steady-state distribution of NEAs by simulating the orbital evolution of test asteroids from the moment they entered the near-Earth region until they reached their respective sinks. The two main outcomes are an outward ejection after a close encounter with, typically, Jupiter or an inward `plunge' towards the Sun. However, for their model to accurately match observations of the NEA population, a complete disruption of NEAs at non-trivial distances from the Sun must be introduced \citep{Granvik2016}. The physical mechanism destroying bodies that reach small-enough perihelion distances is not yet fully understood, but it is believed to be thermal in nature \citep{2021Icar..36614535M} although non-thermal effects have also been proposed \citep{Wiegert2020}. To identify potential mechanisms and guide experimental work to test hypotheses \citep{2022P&SS..21705490T,2021PSJ.....2..165M}, it is crucial to systematically investigate the processes that increase the eccentricity of NEAs close to unity as well as the time-scales involved.

Studies of the dynamical mechanisms that decrease the perihelion distance ($q$) of NEAs have been carried out before, but due to limited computational capabilities, they only considered a relatively small number of test asteroids. \citet{Farinella1994} performed a numerical simulation with 47 test asteroids initially placed close to the $\nu_6$ secular resonance and to comet 2P/Encke. They found that 19 of them were effectively pushed into the Sun by the $\nu_6$ secular resonance and the 3:1 mean-motion resonance (MMR) with Jupiter, aided by secular oscillations of the NEAs' eccentricity ($e$) and inclination ($i$), i.e., the Kozai mechanism.  

\citet{Froeschle1995} extended this study by investigating the effects of additional secular resonances on the evolution of NEAs. They integrated the orbits of 24 known NEAs located close to the main secular resonances between $2<a<3\au$. They separated the integrated asteroids into two categories, the `fast-track', which consider asteroids trapped in resonances, and the `slow-track', which consider asteroids the orbits of which are dominated by close encounters with the terrestrial planets. They verified the importance of $\nu_6$ as a standalone mechanism for reducing $q$, and proposed that $\nu_5$ and second order secular resonances can act as auxiliary mechanisms. Finally, for the first time it was highlighted that $\nu_5$ and $\nu_{16}$ are present at small semimajor axes ($a$).

\citet{Jopek1995} integrated the orbits of 17 bolides for 1 Myr in the past. They found that half of the studied cases belonged to the `fast-track' category -- in the terminology coined by \citet{Froeschle1995} -- in or near a secular resonance or an MMR with Jupiter, and the rest were `slow-track' objects.  The estimated dynamical lifetime of the asteroids was found to be less that 1 Myr.   

Next, \citet{Gladman1997} painted a picture of some typical evolutionary paths followed by NEAs. In their study, they simulated the evolution of a synthetic population of $\sim1500$ members of asteroid families, initially placed near strong MMRs and $\nu_6$. They found that the 3:1 and 5:2 MMRs with Jupiter and $\nu_6$ are more efficient in delivering NEAs to orbits very close to the Sun. In a follow-up study, \citet{Gladman2000} integrated the orbits of 117 known NEAs. They underlined the importance of other resonances in driving NEAs close to the Sun in addition to the 3:1 MMR and $\nu_6$, such as $\nu_2$ and $\nu_5$ at low $a$. They also noted that $\nu_3$ and $\nu_4$ are able to increase the $e$ of NEAs, making their orbits Earth-crossing and, thus, facilitating close encounters.

\citet{Foschini2000} expanded the study of \citet{Jopek1995}, and integrated the orbits of 20 bright bolides for a time span of at least 10 Myr. They reported the same mechanisms for reducing $q$ as \citet{Farinella1994} and \citet{Gladman2000}, stressing the efficiency of overlapping secular resonances.  

\citet{Vokrouhlicky2012} studied asteroid 2004~LG which they propose has recently been very close to the Sun. They simulated its past and future evolution and found that, while being in the 4:1 MMR with Jupiter, the secular oscillations of the Kozai mechanism have already brought it as close as $0.026\au$ from the Sun $3000\yr$ ago and it will likely fall into the Sun within the next $10^4\yr$. In addition, \citet{Pichierri2017} developed a semi-analytic model for the restricted planar three-body problem and found that the 4:1 MMR with a planet with $e\leq0.1$ reduces the $q$ of an asteroid, in an initially almost circular orbit, more efficiently compared to other MMRs. \citet{Emel'yanenko2017} investigated the short-term evolution of known asteroids due to the Kozai secular variations of their $e$ and $i$ in an effort to identify asteroids that have recently been close to the Sun. To that end, he integrated the orbits of $100$ clones of known asteroids with a small value of the vertical component of orbital angular momentum and found 11 asteroids that have reached $q<0.1\au$ at some point in the past $10^4\yr$.

Due to computational limitations, \citet{Farinella1994,Froeschle1995,Gladman1997,Gladman2000,Foschini2000} have mostly relied on small samples of orbits of NEAs and identified resonances by eye inspection of the resonant arguments. Our aim is to extend these studies and examine most of the possible dynamical mechanisms that reduce $q$ to the average disruption distance $q^*=0.076\au$ \citep{Granvik2016} by using the dataset generated from the dynamical simulations carried out by \citet{Granvik2017,Granvik2018} that we have at our disposal. Due to the large number of test asteroids, the identification of mechanisms that decrease $q$ cannot be done by visual inspection. Hence, an automated algorithm must be developed. In this study, we will also determine how these different mechanisms relate to the dynamical lifetime of NEAs at small $q$, that is, the time-scale over which the orbits of NEAs are pushed close to the Sun, as well as the relevant time spent by NEAs at different heliocentric distances. In addition, we aim to determine whether the last recorded mechanism has been the most efficient in driving the asteroid close to the Sun, and if not, determine which one has been the most efficient during orbital evolution of the asteroid. Finally, we will measure the fraction of cases for which the mechanism that brought the asteroid from the asteroid belt into the near-Earth region, coincided with the mechanism that drove it close to the Sun.

\section{Theory and methods}\label{sec:methods}

\subsection{Lifetime of asteroids with small perihelion distances}

\citet{Granvik2017,Granvik2018} performed extensive simulations of the orbital evolution of a synthetic population of asteroids originating in the asteroid belt and their subsequent evolution in the near-Earth region. For the orbital integrations, they used the SWIFT RMVS4 integrator \citep{LevisonDuncan1994} with a 12-hour time-step. RMVS4 can handle planetary encounters and was tailored to accommodate the needs of the project. The simulations provide us with an appropriate sample of test asteroids. We focus on the part of the evolution of the test asteroids from the moment they escape the asteroid belt and become NEAs ($q<1.3\au$), until they reach their respective sinks; typically a collision with the Sun or an ejection from the inner regions of the Solar System.  

In our analysis, we want to take into account the super-catastrophic disruption of asteroids near the Sun, and, consequently, we consider test asteroids that acquire a $q$ lower than the average disruption distance $q^*$ completely disrupted, and disregard any future evolution. Test asteroids for which $q$ never becomes small enough are discarded. While an instantaneous total destruction of asteroids at $q=q^*$ is not a realistic assumption, it is a simplification that is required for the purposes of this study, since the exact destruction mechanism is still under investigation.

We define the dynamical lifetime of NEAs with small $q$, $\tau_{lq}$, to be  the total amount of time it takes for their $q$ to go from $q_l=0.4\au$ to $q^*$. We start counting at the first occasion that the $q$ of a test asteroid reaches below $q_l$, although, as $q$ oscillates, $q$ can still rise above $q_l$. The value of $q_l=0.4\au$ was chosen as the upper limit because it is the approximate heliocentric distance of Mercury and, in addition, there is an absence of sub-meter-sized boulders with $q\lesssim0.4\au$, suggesting that objects with diameters $D\lesssim1\meter$ are destroyed and ground down to mm-scale and smaller fragments inside of the orbit of Mercury \citep{Wiegert2020}.  

\subsection{Time spent at different distances close to the Sun}

\subsubsection{Bin size}\label{subsec:bins}

\begin{figure}
\centering
\includegraphics[width=0.48\textwidth]{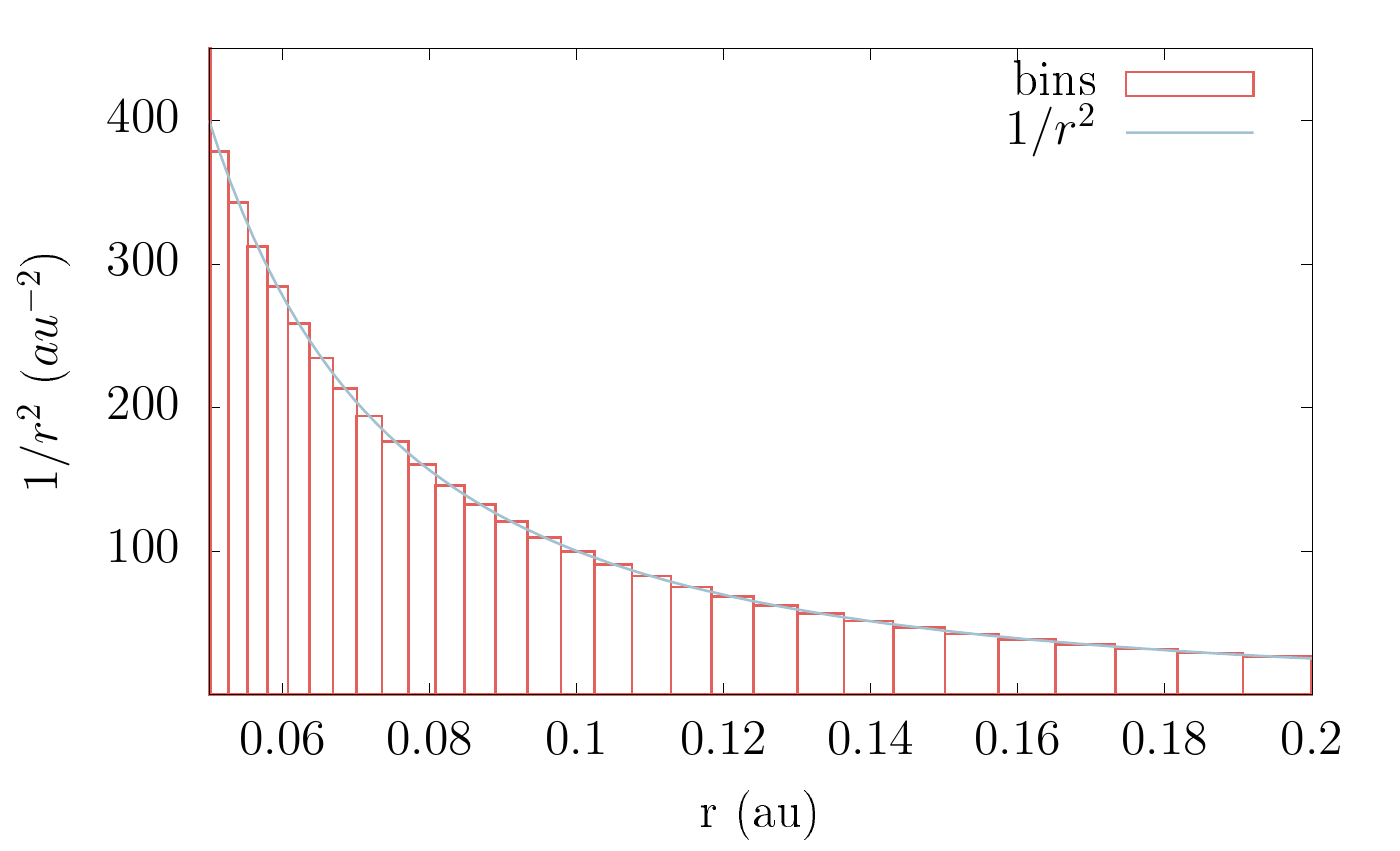}
\caption{The region between $0.05<r<0.2\au$ divided in 29 bins according to $r^{-2}$ with the edges of each bin defined by points with $10\%$ difference in $r^{-2}$. The goal is to study the time each NEA spends at different distances from the Sun. In addition, we split the range between $0.2\au<r<0.4\au$ in four bins of width $0.05\au$ and also include another bin for $r<0.0502\au$, for a total of 34 bins.}
\label{fig:rbins}
\end{figure}

NEAs spend a different amount of time at different heliocentric distances, and we are mostly interested in the amount of time spent in the vicinity of the Sun. The first step for measuring this time is splitting $q \leq q_l$ in separate intervals, or bins. Although we are here mostly interested in $q>q^*$, we will extend the $q$ range all the way to the solar radius, which is the heliocentric distance at which an asteroid would collide with the solar photosphere and therefore a condition that, when met, stops the orbital integration of a test asteroid. The range $0.2\au<q<0.4\au$ is divided in four bins of width $0.05\au$. The region below $0.2\au$ is divided according to $r^{-2}$, where $r$ is the heliocentric distance, since the irradiation from the Sun is believed to be responsible for the disruption of asteroids with small $q$, and the irradiation is proportional to $r^{-2}$. The edges of the bins are defined by points each one of which has a 10\% difference in $r^{-2}$ compared to the preceding one, starting from the inside out. This leads to 29 additional bins, with the innermost bin's edge located at $0.0502\au$ (Fig.~\ref{fig:rbins}). Additionally, we include a bin that covers the area below $0.0502\au$, for a total of 34 bins. Our aim is to determine the time that NEAs stay within a certain interval in i) $q$ and ii) $r$.  

\subsubsection{Distribution in perihelion distance $q$} 

During the dynamical evolution of each test asteroid that eventually reached $q<q^*$, we record every occasion that $q$ enters one of the $q$ bins. In the simulations, the orbital elements were recorded every $250\yr$. As a result, the total time that an asteroid has a $q$ within a given $q$ bin is found by multiplying the number of recorded occasions by $250\yr$. Adding up every interval of time recorded in every bin, gives a result that is not equal to the total lifetime $\tau_{lq}$, but will typically be smaller. The reason is that $q$ can become larger than $q_l$ during its evolution, and these occasions are not registered, since these values lie outside of the total interval covered by the $q$ bins.

\subsubsection{Distribution in heliocentric distance $r$}

Similarly to the distribution of the time spent at different $q$, we also compute the total time each test asteroid spends having a heliocentric distance $r$ that falls within an $r$ bin. Note that the boundaries of the $r$ and $q$ bins are the same (see Sec.~\ref{subsec:bins}).

First, we need to calculate the time an asteroid spends inside shells of radii defined by the edges of the $r$ bins. The total time a test asteroid orbits inside the radius of a specific shell, is found as follows. From the equation of the ellipse
\begin{equation}\label{eq:ellipse}
r=\frac{a\left(1-e^2\right)}{1+e \cos{f}}\,,
\end{equation}
where $f$ is the true anomaly at each output time-step, we calculate the true anomaly at the radius of a shell $r=r_0$
\begin{equation}
f_0=\arccos{\left( \frac{1}{e}\left(\frac{a\left(1-e^2\right)}{r_0}-1\right)\right)}\,.
\end{equation}
Then, the area of the ellipse for $r<r_0$ is found by 
\begin{equation}
A=\frac{1}{2} \int_{-f_0}^{f_0} r^2 \,df\,.
\end{equation}
 By first dividing the resulting area with the total area of the ellipse and then multiplying with $250\yr$, we get the total amount of spent inside $r_0$ at each output time-step. Finally, to get the time spent in each $r$ bin, we subtract the time spent inside the cell corresponding to the inner edge of the bin, from the time below the adjacent outer shell, i.e., the outer edge of the bin.  

\subsection{Resonant mechanisms increasing the eccentricity of NEAs}\label{sec:lib_mech}

In what follows, we give an overview of mechanisms that are capable of increasing the $e$ of NEAs, because the main objective of this study is to determine how efficient various resonant mechanisms are at reducing their perihelion distances $q=a(1-e)$. On the other hand, close encounters with planets can, in principle, have effects in the $a$ and $e$ of a test asteroid that range from negligible to grave. However, the integrator used for the Granvik et al. NEO model simulations does not keep track of close encounters. Consequently, an extensive study is required to identify, categorise, and scale the frequency of close encounters between planets and NEAs, and measure the effect on the orbits of the latter.   Thus, in this study, we focus solely on resonant phenomena.

The motion of an NEA in the test-particle approximation can be described by the Lidov-Kozai mechanism \citep{Zeipel1910,Kozai1962,Lidov1962}. Considering a perturber (Jupiter) in a circular orbit, one can truncate the Hamiltonian at a quadrupole level (keeping second order terms) in the ratio $a_\text{ast}/a_\text{J}$, and averaging this term once over the orbital motion of the planet and another time over the motion of the asteroid. The resulting Hamiltonian is then independent of the longitude of the ascending node, $\Omega$. Consequently, the vertical component of the angular momentum $J_z$ and the total energy are constant, and the system is integrable. 

For an NEA with small $e$ and $i$, the argument of perihelion, $\omega$, circulates over $360^{\circ}$, and $e$ and $i$ can be considered constant. For larger values of $i$, $\omega$ still circulates, but at the same time $e$ oscillates coupled with $i$. At $\omega=90\degree$ or $\omega=270\degree$, $e$ assumes a maximum and $i$ a minimum value. Above a critical threshold value of $i_\text{max}\simeq39.2\degree$ (and below $140.8\degree$), a separatrix is created at $e=0$ that divides the phase space in regions where $\omega$ can either circulate or librate around $90\degree$ or $270\degree$. For increasing $i$, the libration regions become larger, so the resonance becomes stronger. Also, as $e=0$ is an unstable equilibrium point, asteroids with large $i$, even if they initially have small $e$, will be forced to acquire large maximum $e$ values while oscillating \citep{Morby2002book}.

The quadrupole approximation can be useful in gaining insight about the evolution of many NEAs. However, for certain NEAs, this approximation is inefficient in reproducing their exact orbits, due to Jupiter's non-zero eccentricity $e_J=0.0489$. Considering the effects of an eccentric perturber, \citet{Ford2000,Lithwick2011,Katz2011,Naoz2011,Naoz2012,Naoz2013} studied the Kozai mechanism in the octupole approximation, keeping an additional, third-order term in the Hamiltonian. The new form of the mechanism is called the eccentric Kozai-Lidov (EKL) mechanism. In this case, $J_z$ is not conserved, and the orbit of the asteroid is characterized by large excitations of $e_\text{ast}$, large jumps in $i_\text{ast}$ that lead to `flips' in the direction of their motion from prograde to retrograde, and vice versa, as well as chaotic motion. During the `flips', $e_\text{ast}\rightarrow1$ as $i_\text{ast}\rightarrow90\degree$. The relative size of the octupole-order term of the Hamiltonian compared to the quadrupole-order term is measured by
\begin{equation}\label{eq:epsilon}
    \epsilon=\frac{e_J}{1-e_J^2}\frac{a_\text{ast}}{a_J}\,.
\end{equation}
\citet{Antognini2015} estimated that the time-scale of the EKL variations $\tau_\text{EKL}$ scale with the time-scale associated with the classical Lidov-Kozai mechanism $\tau_\text{LK}$ as $\tau_\text{EKL}=\frac{\tau_\text{LK}}{\sqrt{\epsilon}}$. Hence, given enough time, the $e$ of an NEA with a relatively large $a$  can increase and become large enough that $q<q^*_\text{dis}$. However, the time-scale of reducing $q$ can be affected by the presence of MMRs and secular resonances.

An MMR occurs when there is a commensurability between the orbital period of two objects, in our case a planet and an NEA. The ratio between their mean motions can have the form
\begin{equation}
\frac{n_\text{ast}}{n_\text{pl}}=\frac{P+Q}{P}\,,
\end{equation}
where $P$ and $Q$ are positive integers. $Q$ also defines the order of the resonance. In this study we will only consider MMRs with Jupiter and the Earth. We also tested MMRs with Venus, but found that very few objects were caught in them, and therefore decided to ignore them.

Secular resonances occur when the free precession rate of longitude of perihelion, $\varpi$ (or $\Omega$) of an asteroid is equal to the $g$ (or $s$) eigenfrequency of the Solar System, or a combination of those. Here, we only consider resonances associated with $\dot{\varpi}$, since they are the ones that affect the $e$ of an asteroid. Typically, a secular resonance is labelled by $\nu_\text{pl}$ when $\dot{\varpi}=g_\text{pl}$. Note that the $e$ of an asteroid that is captured in a secular resonance can either increase or decrease, depending on its secular phase. 

\subsubsection{Identifying librations in the resonant arguments}

In what follows, we describe an algorithm that identifies librations in the resonant argument of mean-motion and secular resonances during the orbital evolution of the test asteroids in the near-Earth region. Assuming that the orbits of an NEA and a planet are circular and co-planar, the resonant argument of an MMR is defined as:
\begin{equation}
\phi_\text{MMR}=(P+Q)\lambda_\text{pl}-P\lambda_\text{ast}-Q\varpi_\text{ast}\,,
\end{equation}
where $\lambda$ is the mean orbital longitude, which is equal to $\lambda=\varpi+M$, and $M$ is the mean anomaly. However, in the elliptic case, the resonant argument also consists of $\varpi_\text{pl}$, $\Omega_\text{pl}$ and $\Omega_\text{ast}$ terms and the MMR has the structure of a `resonant multiplet'. For the scope of this study, it is enough to restrict ourselves to the use of the principal resonant argument, because the secular frequencies $\dot{\varpi}$ and $\dot{\Omega}$ are small compared to the orbital frequencies $\dot{\lambda}$.

We consider the following MMRs: 2:3E, 1:2E, 5:1J, 4:1J, 7:2J, 3:1J, 8:3J, 5:2J, 7:3J, 9:4J, 11:5J, 2:1J, 5:3J, 3:2J, where E stands for Earth and J stands for Jupiter. For each test asteroid, we only take into account the resonance that lies closest with respect to its $a_\text{ast}$. To find out if an asteroid is captured in an MMR, we look for librations in the resonant argument around $180^\circ$ or $0^\circ$, and also require that the average value of $a_\text{ast}$ during the considered timespan, falls within twice the width of the MMR, as calculated in \citet{Gallardo2021}. 

Identifying secular resonances requires another approach. This is done by detecting a libration in the angle
\begin{equation}
\phi_\text{SR}=\varpi_\text{ast}-\varpi_\text{pl}\,.
\end{equation}
However, we do not take the modulo of this argument with $360\degree$, but, instead, we let the angle rise above $360\degree$ and continue accumulating, cycle after cycle. Then, we perform a linear least squares fit on the time evolution of the respective resonant argument. When the slope becomes (close to) zero, we consider a secular resonance to occur. 

We do not identify secular resonances by inspecting the resonant argument $\varpi_\text{ast}-g_i t$, where $i=2,...,6$, because of the way the simulations of the synthetic NEA population have been structured. Specifically, they have been divided in subgroups of separate simulations consisting of the giant planets and 50 test asteroids, with different integration times that could last up to several Gyr. As a result, there might not be one single computed value for each eigenfrequency, $g_i$, of the Solar System, although, in practice, the differences between the frequencies calculated for every subgroup would be very small.

However, an additional check is introduced, in which we demand that the $g$ of the test asteroid is not very different from the respective $g_i$ eigenfrequency. We measure the $2^{\text{nd}}-6^{\text{th}}$ eigenfrequencies of the Solar System for a few subsets of simulations and define 'typical values'. Then, we calculate $\dot{\varpi}_{ast}$ using a linear least-squares fit in the time evolution of $\varpi$, not limited to the range $[0,360\degree]$. For each secular resonance, we consider the test asteroid as being in the resonance if the calculated $g$ frequency is $g_\text{i}\pm3\arcsecyr$ for $\nu_5$ and $\nu_6$, and $g_\text{i}\pm1.5\arcsecyr$ for $\nu_2$, $\nu_3$, and $\nu_4$.     

The measured eigenfrequencies of the Solar System can vary from $g_5\simeq4.3\arcsecyr$ to $g_6\simeq28.6\arcsecyr$. So we have to use windows of different sizes to detect librations in the resonant arguments of resonances that have different frequencies. We consider static windows of five different sizes: $7.5\times10^3$, $2.5\times10^4$, $5\times10^4$, $10^5$ and $2\times 10^5\yr$. In addition, we use three windows with dynamically-adjusted sizes. For each individual test asteroid, we determine the window sizes from its recorded $\tau_{lq}$; we choose one window that has a size equal to $\tau_{lq}$, one that is half of $\tau_{lq}$, and one that is double $\tau_{lq}$. If $\tau_{lq}>1.25\times 10^6\yr$, we limit the middle window size to $7.5\times 10^5\yr$, and use half/double of this value for the two other windows.

\subsubsection{Handling of overlapping resonances}

The presence of secular resonances inside MMRs has been studied extensively by \citet{Morby1993} and \citet{Moons1995}. Overlapping resonances is a source of chaotic motion that increases the $e$ of asteroids very efficiently. For the purposes of this study, whenever an MMR is identified alongside a secular resonance, we prioritise the MMR as the main acting resonance. However, we retain the ability to differentiate between cases of MMRs occurring alone or overlapping with secular resonances.

There can also be a synergy between secular resonances in raising the $e$ of test asteroids. In that case, we give priority to $\nu_5$ and $\nu_6$ compared to $\nu_2$, $\nu_3$, and $\nu_4$, because the former are more efficient. In addition, we consider $\nu_3$ and $\nu_4$ as a group and denote them with $\nu_3\nu_4$, since they are overlapping due to the proximity of the $g_3$ and $g_4$ secular frequencies. 

\subsubsection{Performance of the resonance-flagging algorithm}

To test the performance of our resonance-flagging algorithm, we first produce a benchmark sample by identifying the mechanisms responsible for reducing the $q$ of 100 random test asteroids by visual inspection of the time evolution of the resonant arguments of the resonances mentioned above. We then compare the benchmark with the results obtained by applying our resonance-flagging algorithm on the same test asteroids. We find that the resonance-flagging algorithm correctly identifies $\sim95\%$ of the resonances.

\section{Results}

\subsection{Typical dynamical evolution of NEAs}

\begin{figure*}
\centering
\includegraphics[width=0.48\textwidth]{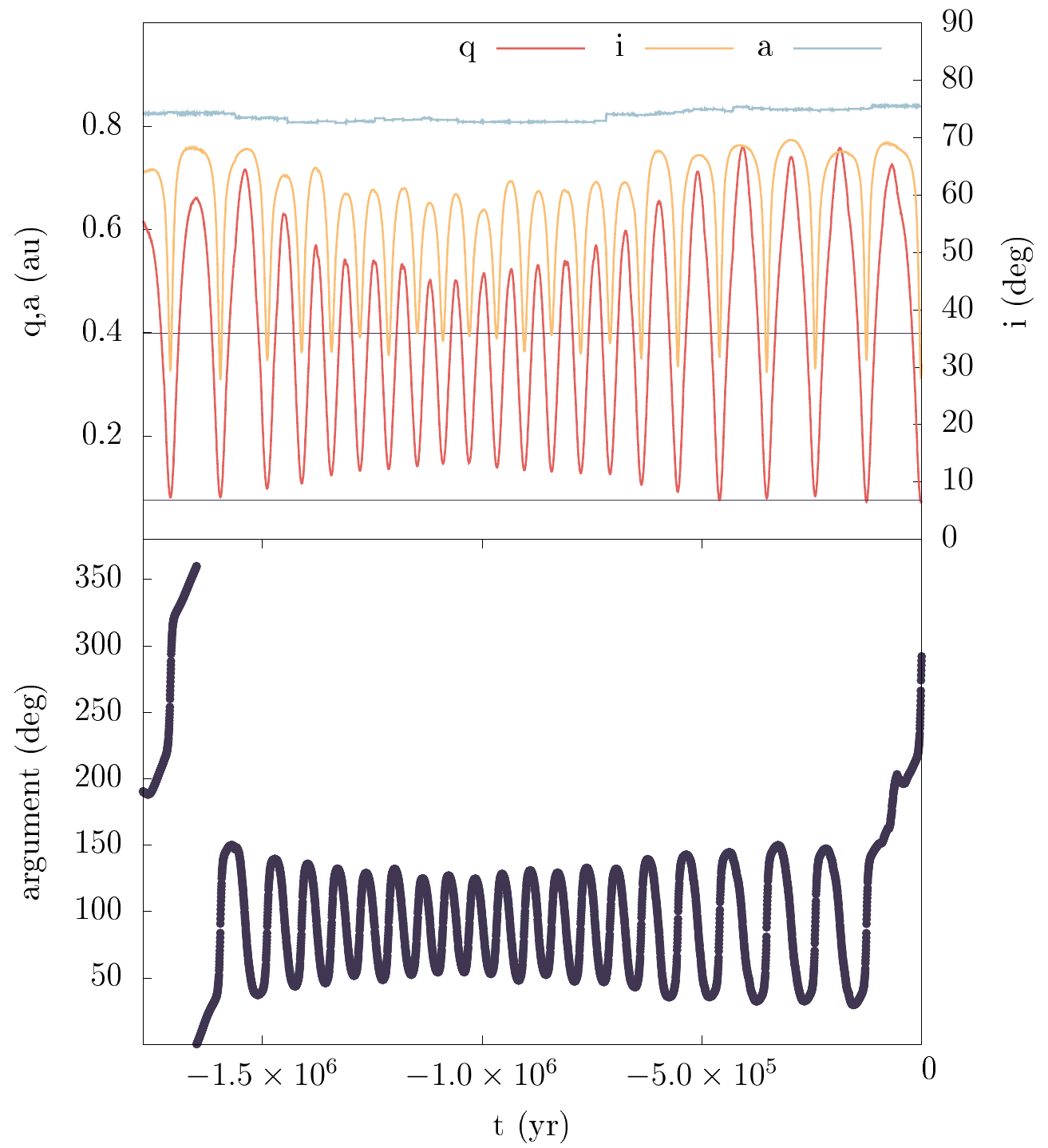}
\includegraphics[width=0.48\textwidth]{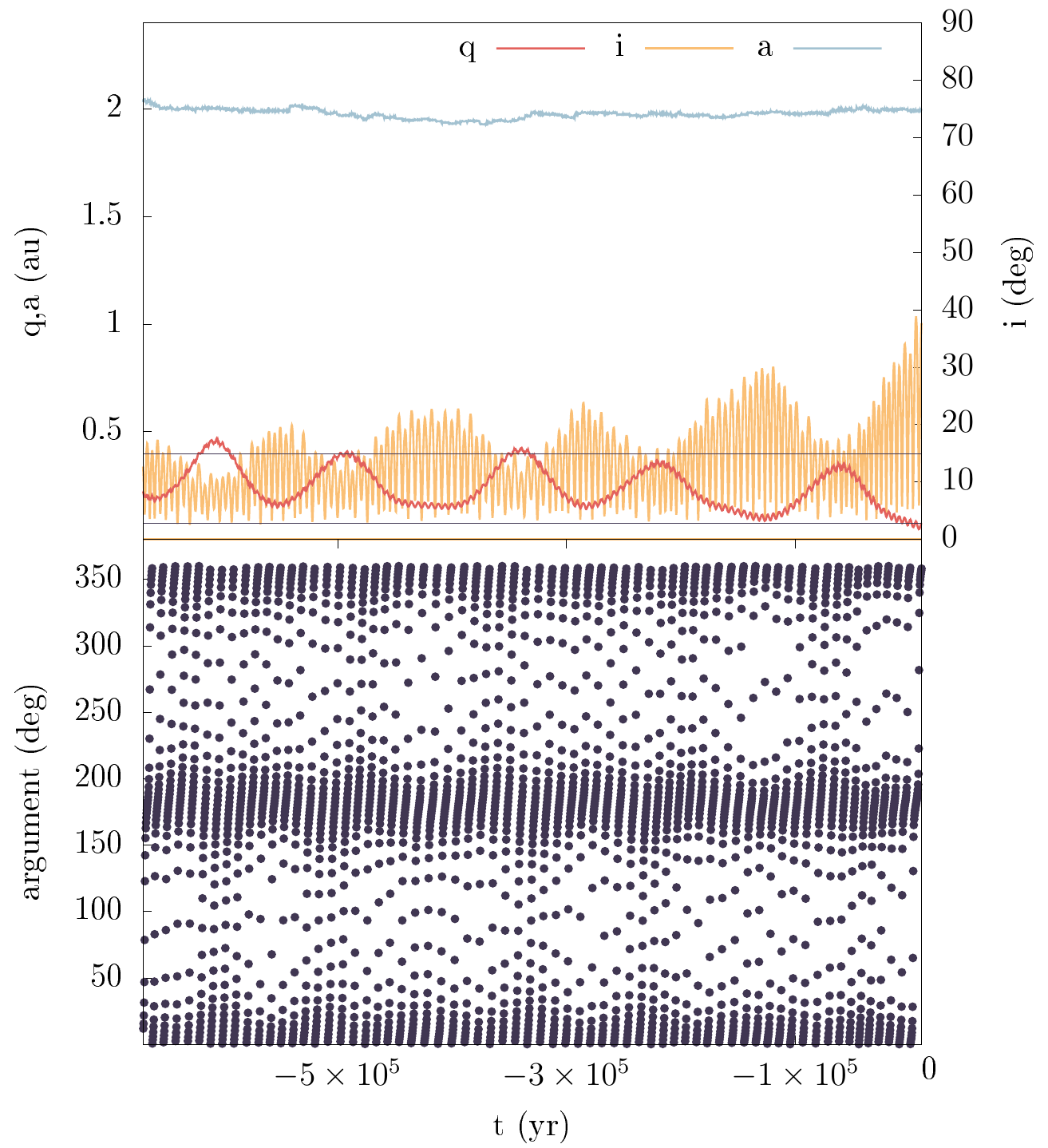}
\caption{Two examples of the final stages of the evolution of test asteroids in the near-Earth region that are subject to the Kozai mechanism. At $t=0$, $q$ becomes lower than $q^*$. Left panels: In the top panel, we show the time evolution of $q$ (red), $i$ (yellow) and $a$ (blue) of a test asteroid that exhibits anti-correlated oscillations between $i$ and $e$. The two horizontal lines correspond to $q_l=0.4\au$, the upper limit under which we start counting $\tau_{lq}$ and $q^*=0.076\au$, the average perihelion distance at which asteroids are destroyed. In the bottom panel, we show the time evolution of $\omega$ for the same test asteroid showing a clear libration around $90\degree$. Right panels: the time evolution of $q$, $i$ and $a$ (top) for a test asteroid that is subject to the Kozai mechanism, but for which $\omega$ circulates.  The $a$ of this asteroid is larger compared to the one shown in the left panels. Thus, the effects of the octupole term in the approximation of the Hamiltonian can be seen here, as the variations of $i$ and $q$ are not bound between a maximum and minimum value, but are increasing.} 
\label{fig:typical_ko}
\end{figure*}

The dynamical evolution of NEAs is the outcome of a complicated interplay between the Kozai mechanism, short-term and long-term trapping into resonances, and close encounters with the planets. In Fig.~\ref{fig:typical_ko}, we show the typical evolution of NEAs that are not trapped in an MMR or a secular resonance during the last stages of their lifetimes. At $t=0$, the $q$ of a test asteroid crosses the $q^*$ threshold, consequently, we present the dynamical evolution prior to this using negative time.

For $\epsilon\ll1$ (defined in Eq.~\ref{eq:epsilon}), that is, for very small $a_\text{ast}$, the motion is dominated by the quadrupole effects. In that case, according to the classical Lidov-Kozai mechanism theory, the oscillations in $e$ and $i$ remain bound. However, in practice,  as we see from the numerical simulations, close encounters between a test asteroid and the planets can change the orbital elements of the former, which in turn affects the amplitude of the $e$ and $i$ variations. As a result, $q$ may pass below the $q^*$ threshold (left panels of Fig.~\ref{fig:typical_ko}). For larger $a_\text{ast}$, and consequently larger $\epsilon$, the octupole term of the approximate hamiltonian starts to affect the motion of NEAs. In the right panels of Fig.~\ref{fig:typical_ko}, we present the case of an asteroid with $\epsilon\sim0.2$ for which the minimum and maximum values of $e$ and $i$ change with each `cycle'. The $e_\text{ast}$ rises much faster compared to the previous very low $\epsilon$ case.  

\begin{figure}
\centering
\includegraphics[width=0.48\textwidth]{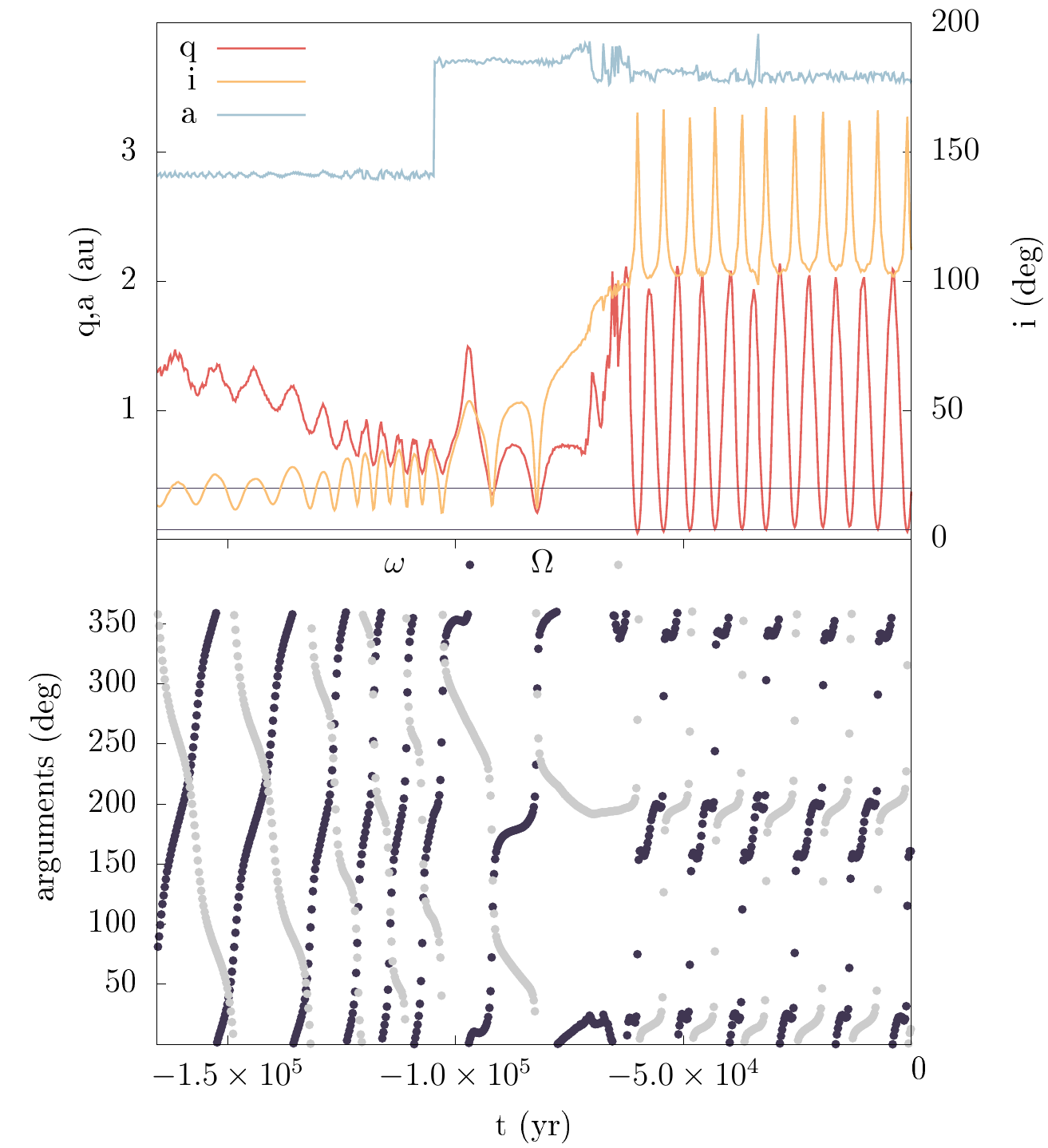}
\caption{Similar to Fig.~\ref{fig:typical_ko}, but for a test asteroid the orbit of which flips, that is, changes from prograde to retrograde, due to the EKL mechanism. Note that, during the flip, both $\omega$(dark blue) and $\Omega$ (grey) librate.} 
\label{fig:flip}
\end{figure}

Most studies of the EKL mechanism are focused on the hierarchical case, in which the perturber's orbit lies sufficiently far from the test particle. \citet{Li2014} argue that, for $\epsilon>0.1$, the hierarchical condition (a tight inner binary, such as the Sun and an asteroid, and a third object in a much larger orbit, such as Jupiter) may break down and one can no longer make accurate predictions based on the theory. However, we can still recognize the EKL effect in the orbital evolution of our simulated test asteroids. In Fig.~\ref{fig:flip}, we show the evolution of a test asteroid with $\epsilon>0.3$ and whose $e_\text{ast}$ gets excited very quickly and also experiences an orbital flip, that is, changes from prograde to retrograde motion, a result in accordance with \citet{Li2021}, that studied the non-hierachical case for EKL. Note that we plot the continued orbital evolution of the test asteroid until it collides with the Sun to illustrate the flip, even though we disregard any evolution after $q_\text{ast}<q^*$ in the subsequent analysis.  

\begin{figure*}
\centering
\includegraphics[width=0.48\textwidth]{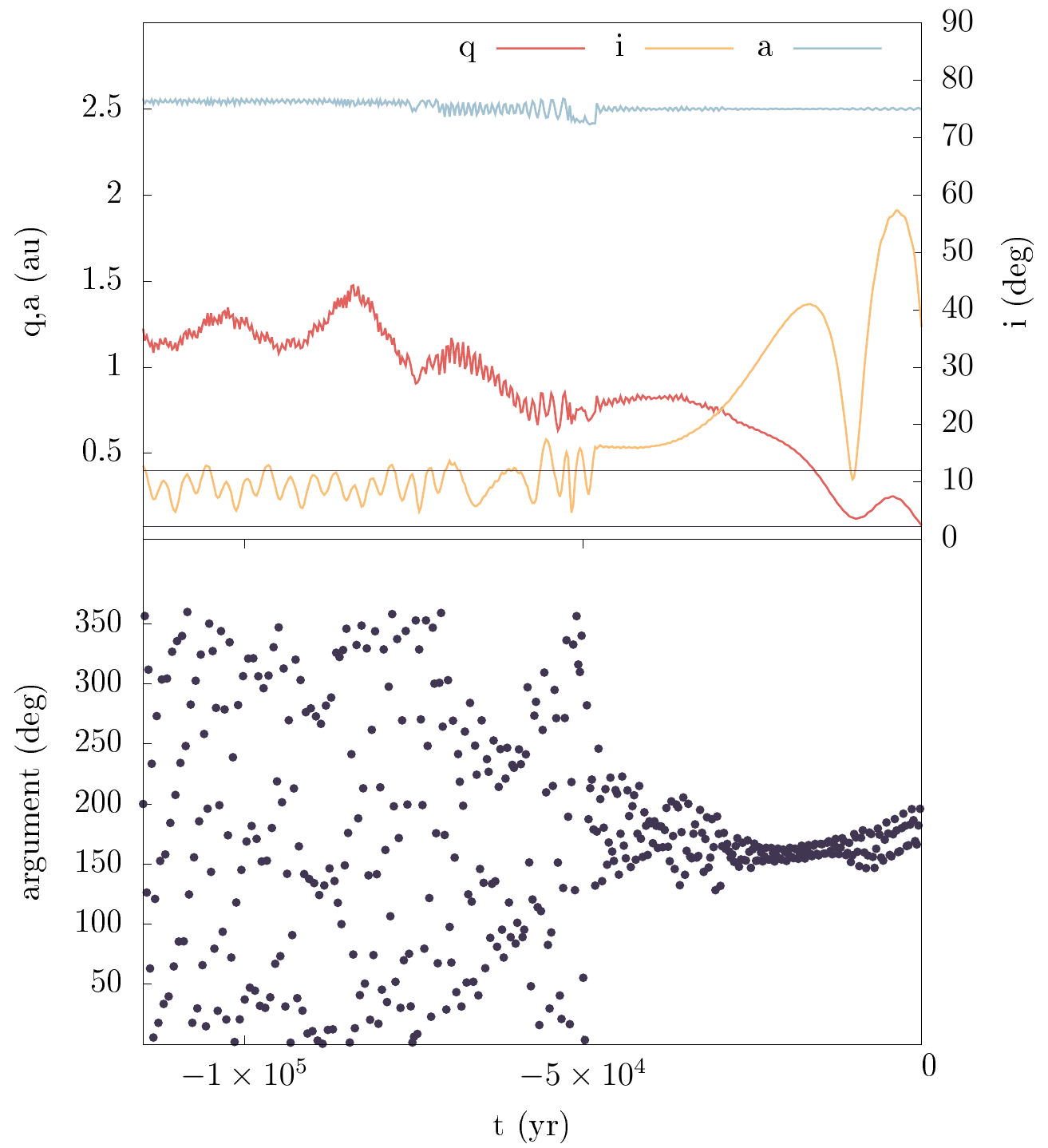}
\includegraphics[width=0.48\textwidth]{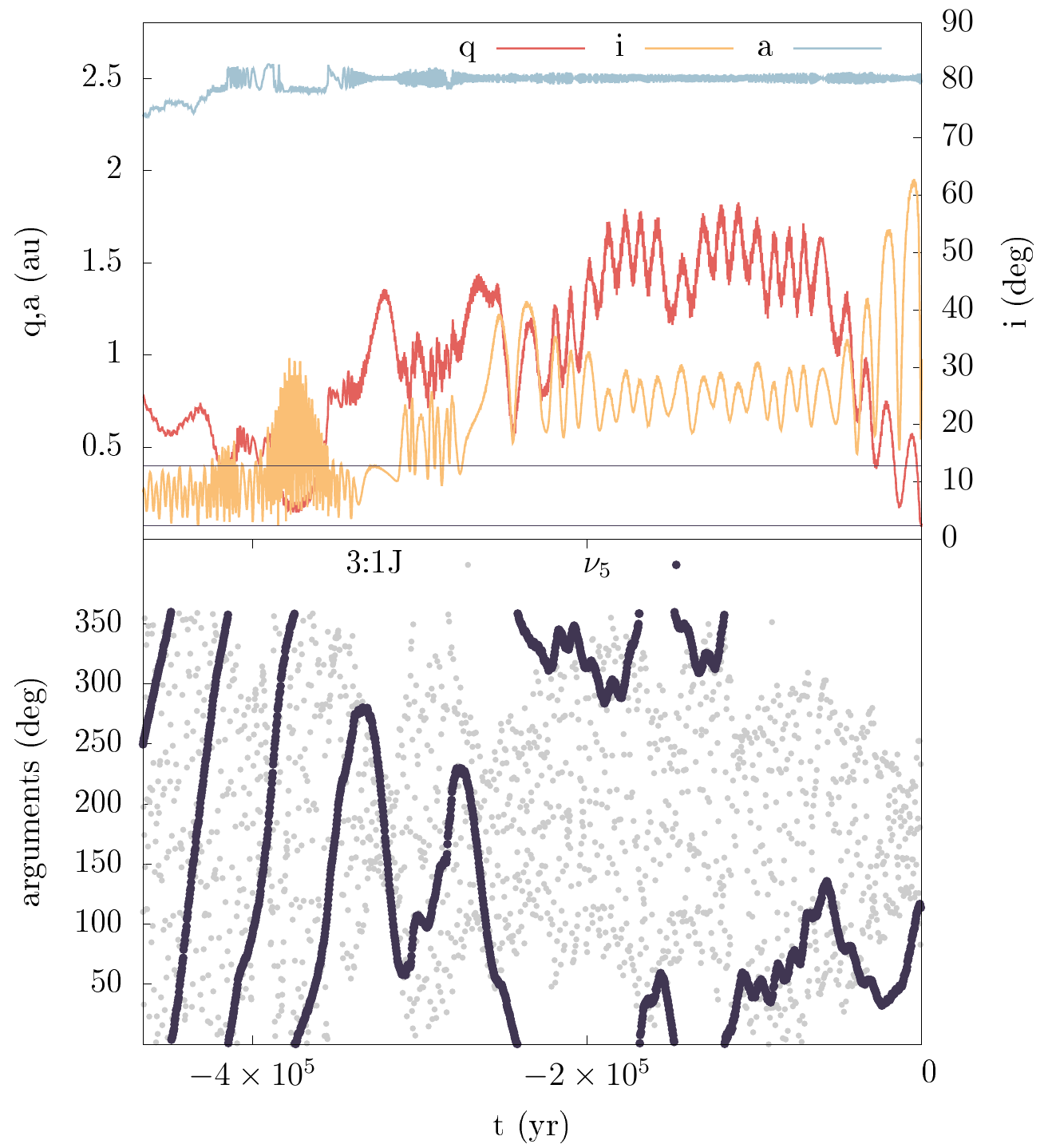}

\includegraphics[width=0.48\textwidth]{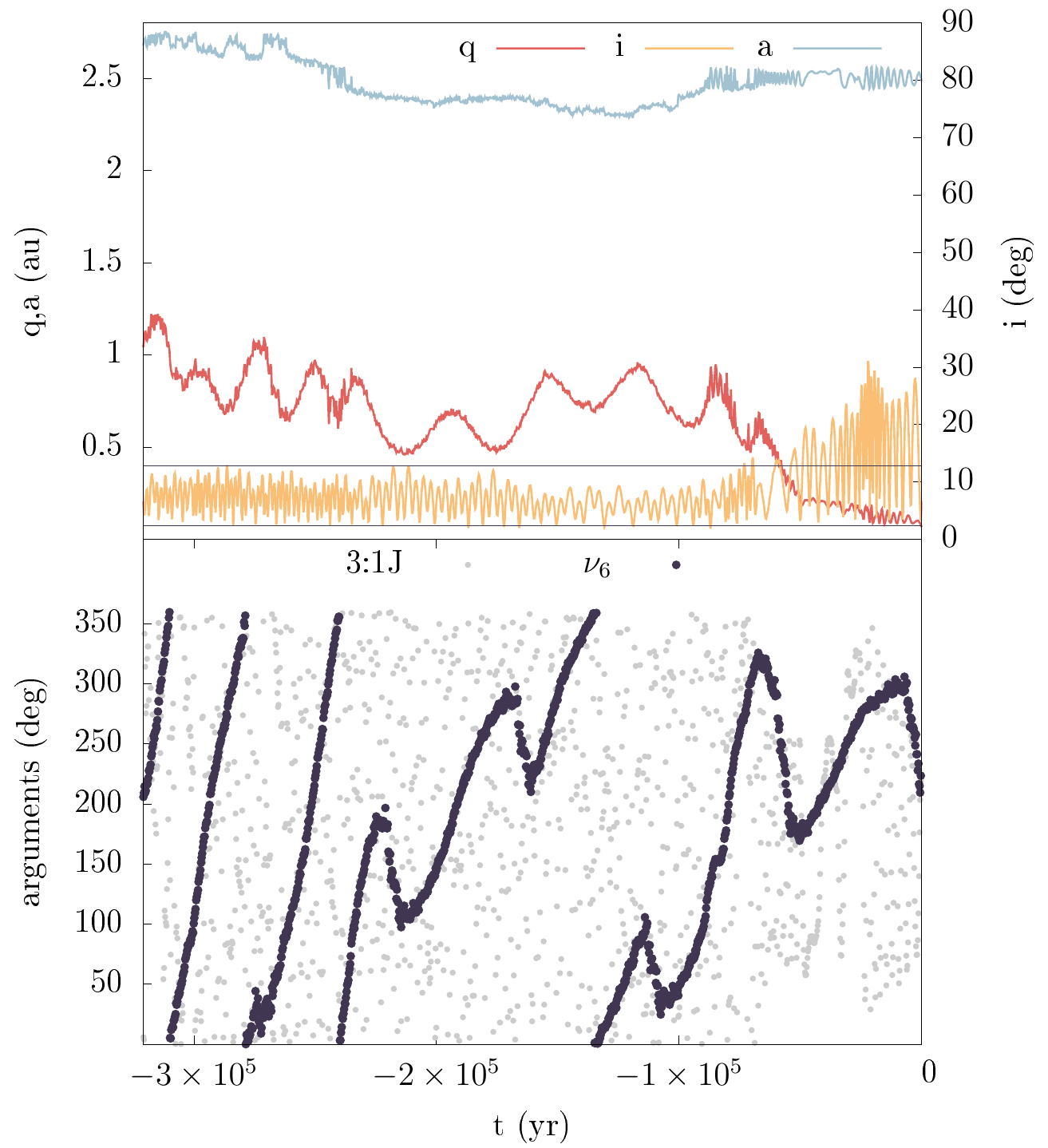}
\includegraphics[width=0.48\textwidth]{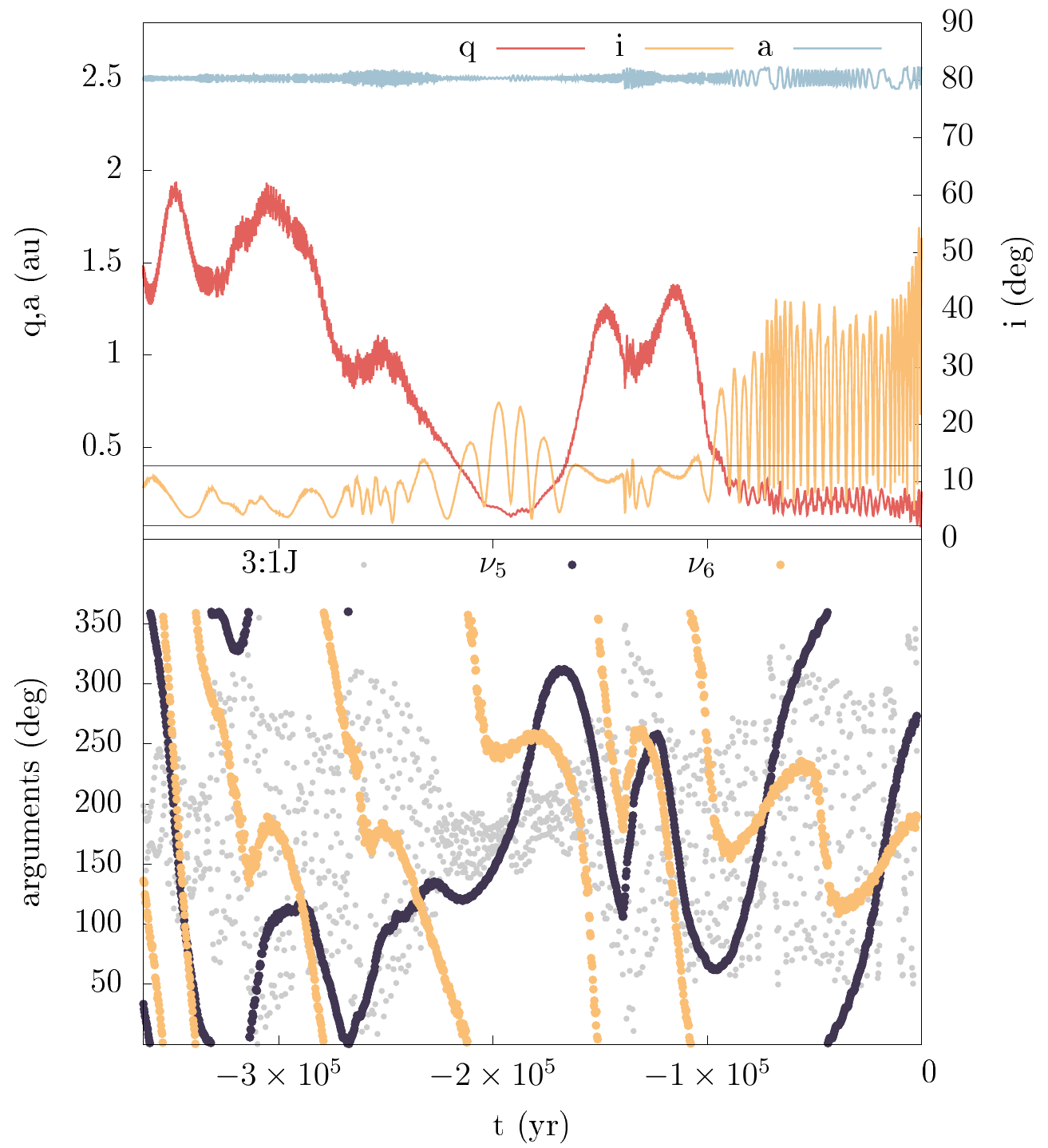}

\caption{Same as Fig.~\ref{fig:typical_ko}, but with examples of four test asteroids that have been captured in an MMR during the last stages of their evolution. Top left panels: In the top plot, we show the time evolution of $q$, $i$, and $a$ for a test asteroid that has been captured in the 3:1J MMR. In the bottom plot, we show the time evolution of the resonant argument of the 3:1J MMR for the same test asteroid, showing a clear libration around $180\degree$. In the three other panels, we show a test asteroid that has been captured in the 3:1J MMR and, additionally, in the $\nu_5$ secular resonance (top right panels), in the $\nu_6$ secular resonance (bottom left panels), and in both the $\nu_5$ and $\nu_6$ secular resonances (bottom right panels). In these plot, the arguments of secular resonances are shown in dark blue (and yellow), and the arguments of the MMRs in grey.}
\label{fig:typical_mmr}
\end{figure*}

A very common dynamical status of NEAs is to be captured in an MMR with a planet. MMRs with Jupiter are the most common and they have the ability to quickly increase the $e$ of NEAs and drive them close to the Sun. In Fig.~\ref{fig:typical_mmr}, we show the last stages of the orbital evolution of a test asteroid that has been captured in the 3:1J MMR, as well as cases where simultaneously with the MMR, a secular resonance ($\nu_5$, $\nu_6$, or both) has been detected by our algorithm.

\begin{figure*}
\centering
\includegraphics[width=0.48\textwidth]{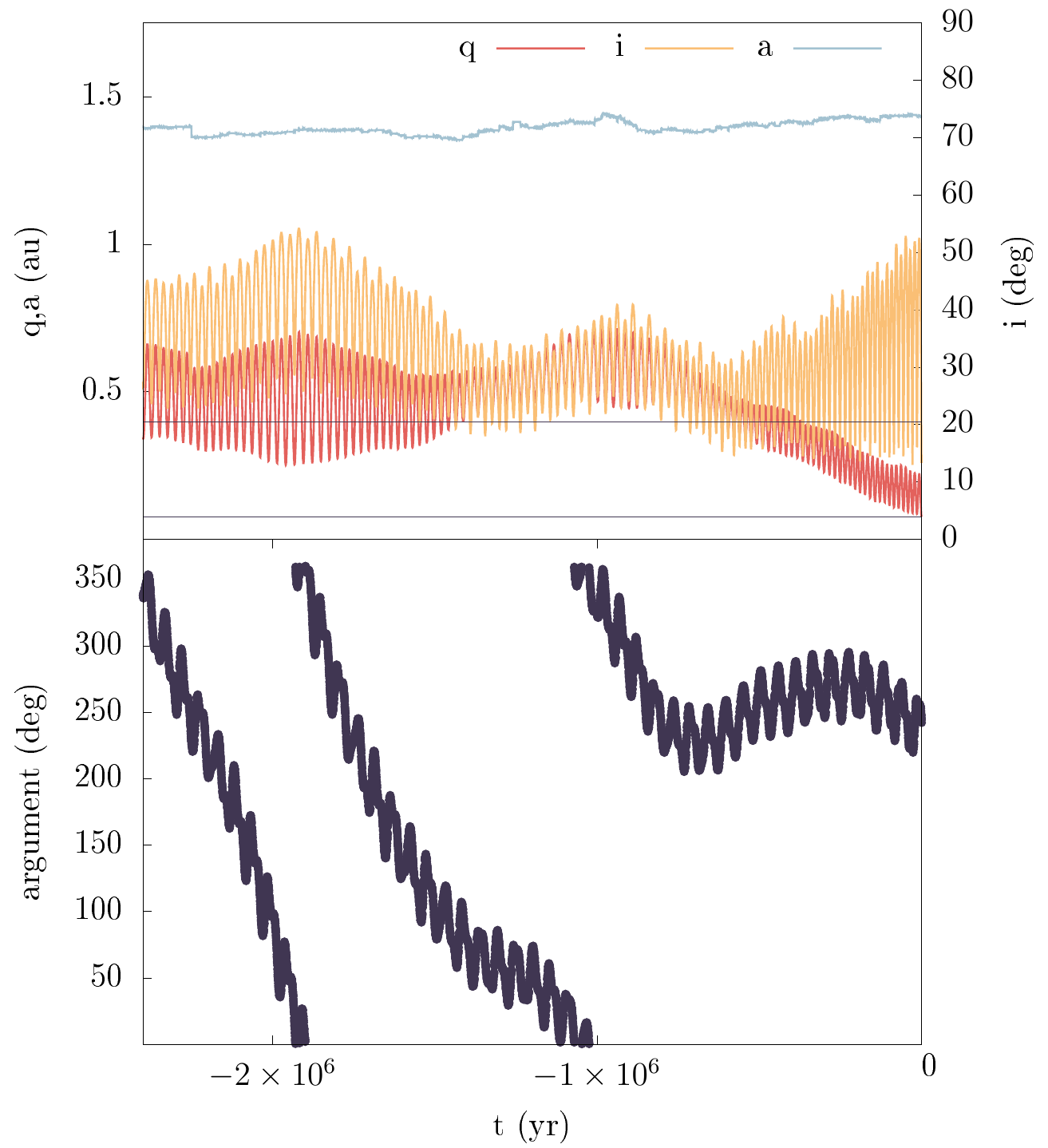}
\includegraphics[width=0.48\textwidth]{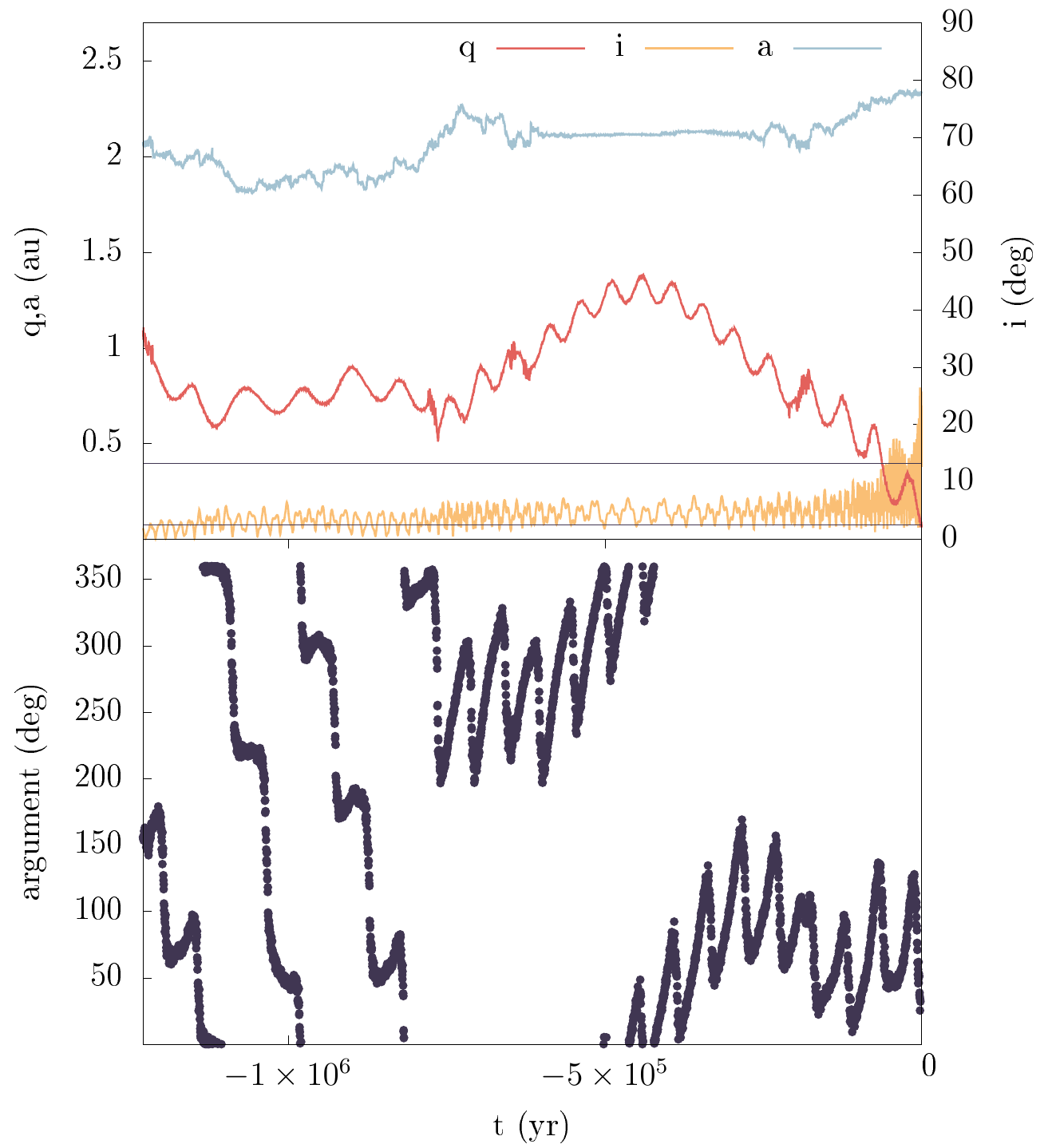}
\caption{Same as Fig.~\ref{fig:typical_ko}, but for a test asteroid that has been captured in the $\nu_5$ secular resonance (left panels) and a test asteroid that captured in the $\nu_6$ secular resonance (right panels). }
\label{fig:typical_v5v6}
\end{figure*}

Another common state in the orbital evolution of NEAs is getting captured in the $\nu_5$ or $\nu_6$ secular resonances, that can also potentially drive their $q$ below $q^*$. Fig.~\ref{fig:typical_v5v6} shows the last stages of the time evolution of the orbital elements of two test asteroids that have been captured in the $\nu_5$ and $\nu_6$ secular resonances.

\begin{figure*}
\centering
\includegraphics[width=0.48\textwidth]{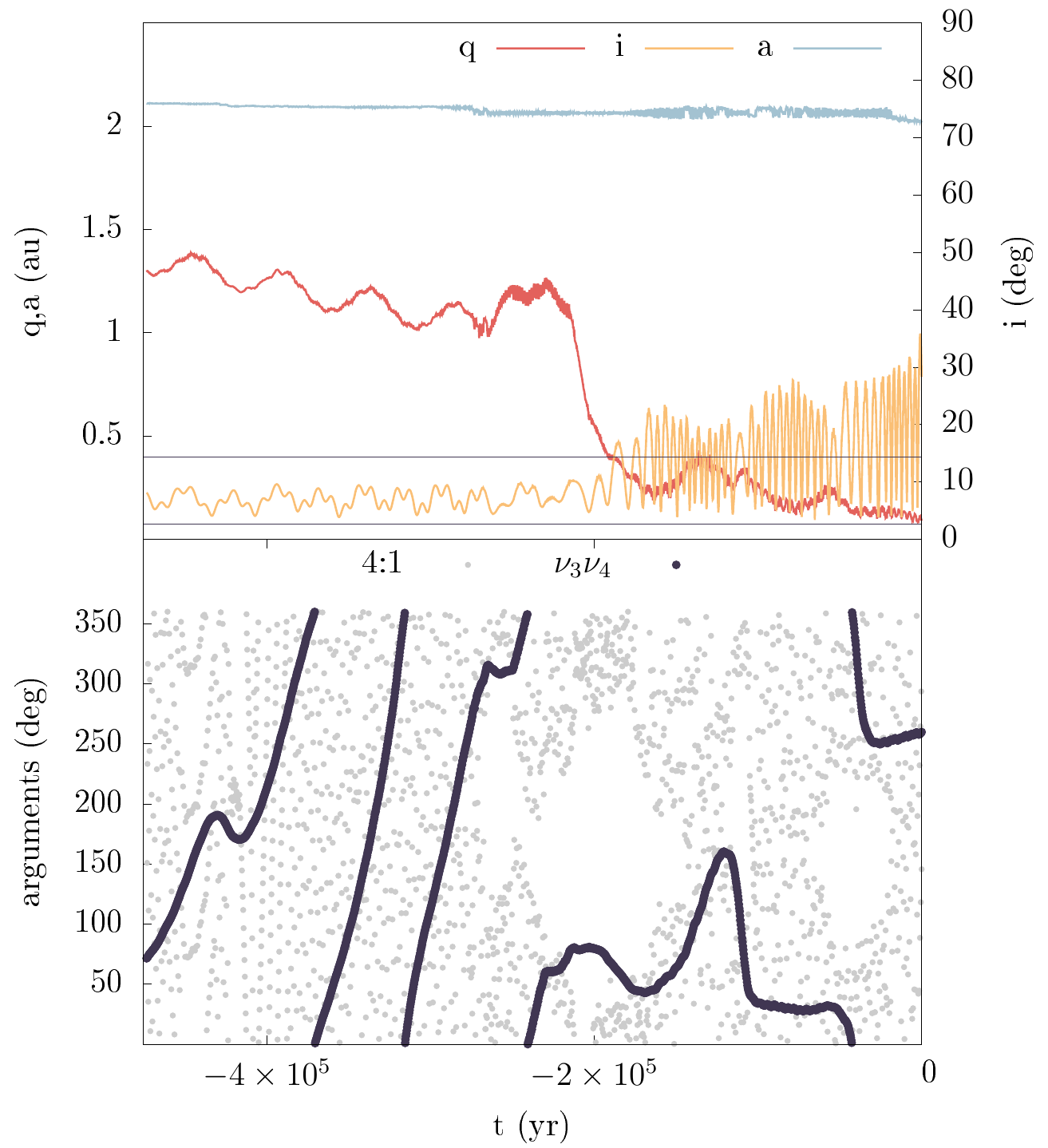}
\includegraphics[width=0.48\textwidth]{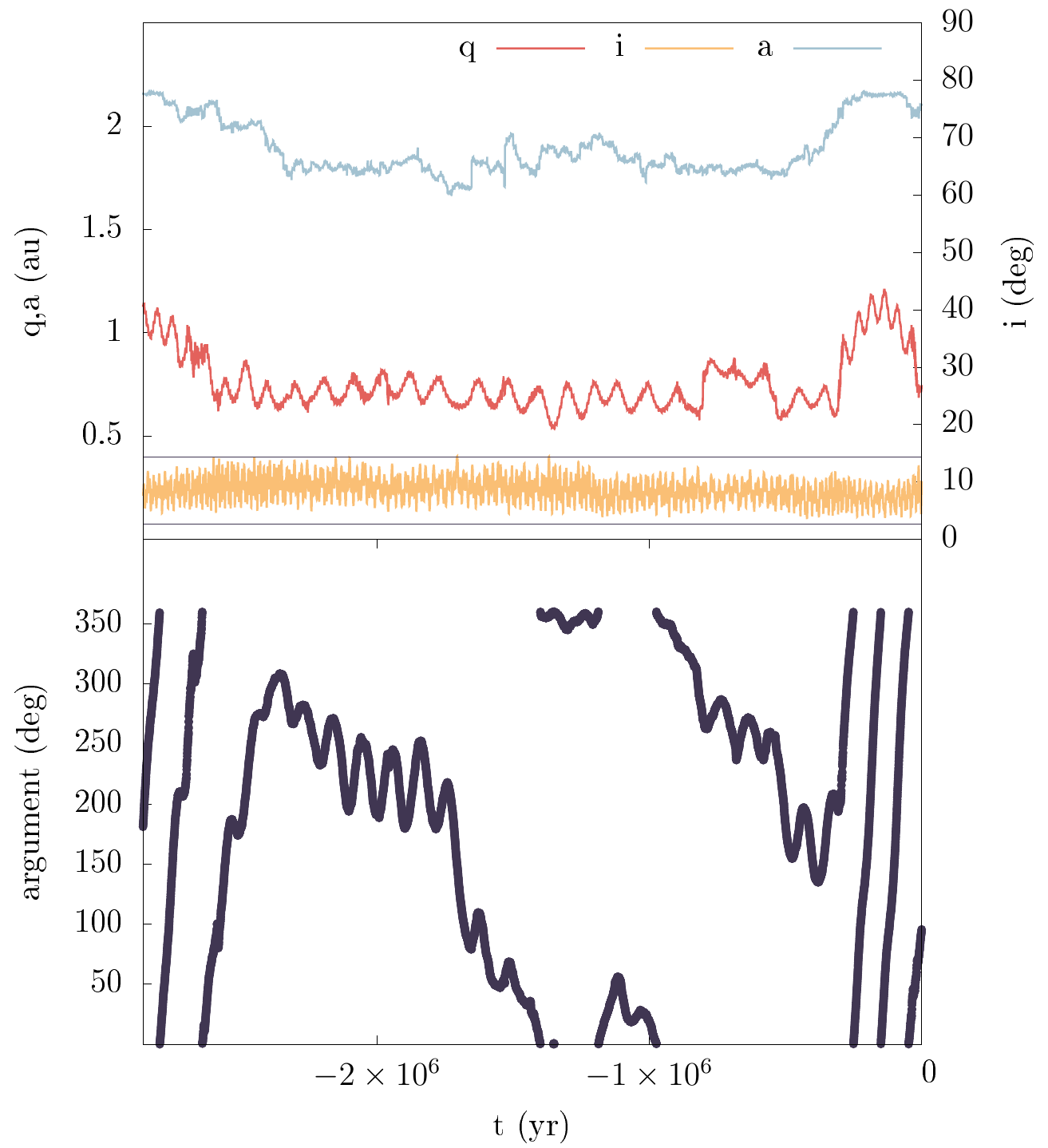}
\caption{Left panels: Same as Fig.~\ref{fig:typical_ko}, but showing typical orbital evolution for a test asteroid that has been captured in the $\nu_3\nu_4$ secular resonance, acting in synergy with 4:1J MMR. Right panels: an example of one part of the orbital evolution of a test asteroid that has been captured in only the $\nu_3\nu_4$ secular resonance. In that case, $t=0$ does not correspond to any particular event but it is adopted for consistency to previous plots and to give an overview of the timescale in which the resonance is acting. }
\label{fig:typical_v3v4}
\end{figure*}

\begin{figure}
\centering
\includegraphics[width=0.48\textwidth]{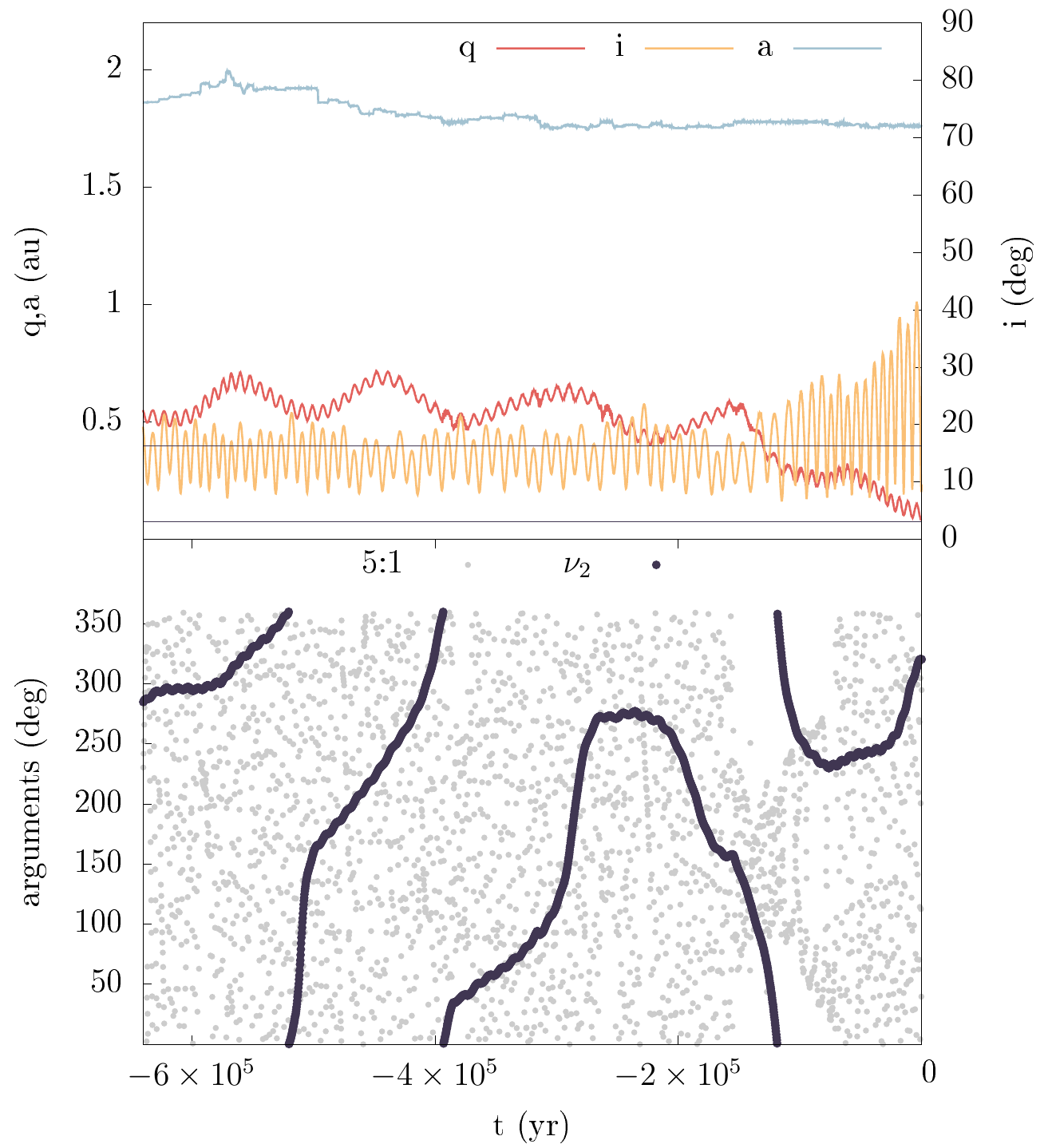}
\caption{Same as Fig.~\ref{fig:typical_ko}, but showing typical orbital evolution for a test asteroid that has been captured in the $\nu_2$ secular resonance.}
\label{fig:typical_v2}
\end{figure}

The $\nu_2$, $\nu_3$, and $\nu_4$ secular resonances are not as strong, consequently, when `acting' alone, they do not affect the average value of $e$ a lot (see Fig.~\ref{fig:typical_v3v4}, right panels). However, by bringing $q$ in the vicinity of the terrestrial planets and by overlapping with other resonances, they can act as auxiliary mechanisms. In Fig.~\ref{fig:typical_v2} we show such a behaviour for $\nu_2$ acting in synergy with the 5:1J MMR, and in Fig.~\ref{fig:typical_v3v4} (left panels) the $\nu_3$ and $\nu_4$ secular resonances, acting in synergy with the 4:1J MMR.

\begin{figure*}
\centering
\includegraphics[width=\textwidth]{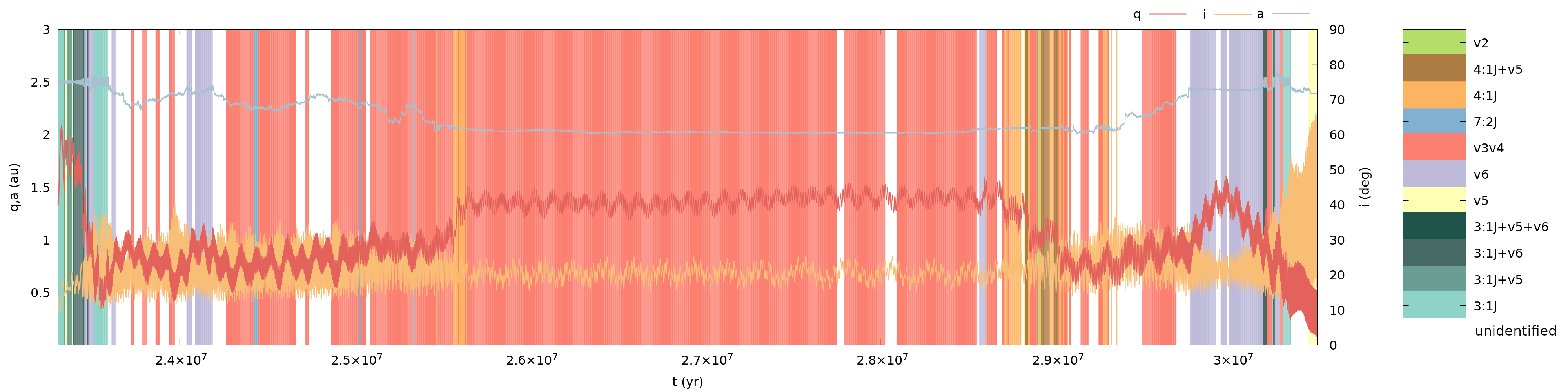}
\caption{An example of the alternation between different resonances occurring during the orbital evolution of an asteroid in the near-Earth region. Here we show the entire time evolution of $q$, $a$ and $i$ of a test asteroid from the moment it becomes an NEA, until it reaches below the average disruption distance $q^*$. The color coding in the background corresponds to the resonances in which the asteroid is captured at that moment.}
\label{fig:example_evolution}
\end{figure*}

The orbit of a typical test asteroid exhibits alternating occurrences of some these resonances, resulting in an intricate evolution with increases and decreases of $e$. In Fig.~\ref{fig:example_evolution} we show the time evolution of $q$, $a$ and $i$ of a typical test asteroid, which starts trapped in 3:1J MMR, switches between a variety of resonances, and ends up crossing $q^*$ trapped in $\nu_5$. The occurrence of each resonance is illustrated with different colours.

\subsection{Orbital resonance just prior to disruption}
\label{sec:last}

\begin{figure}
\centering
\includegraphics[width=0.48\textwidth]{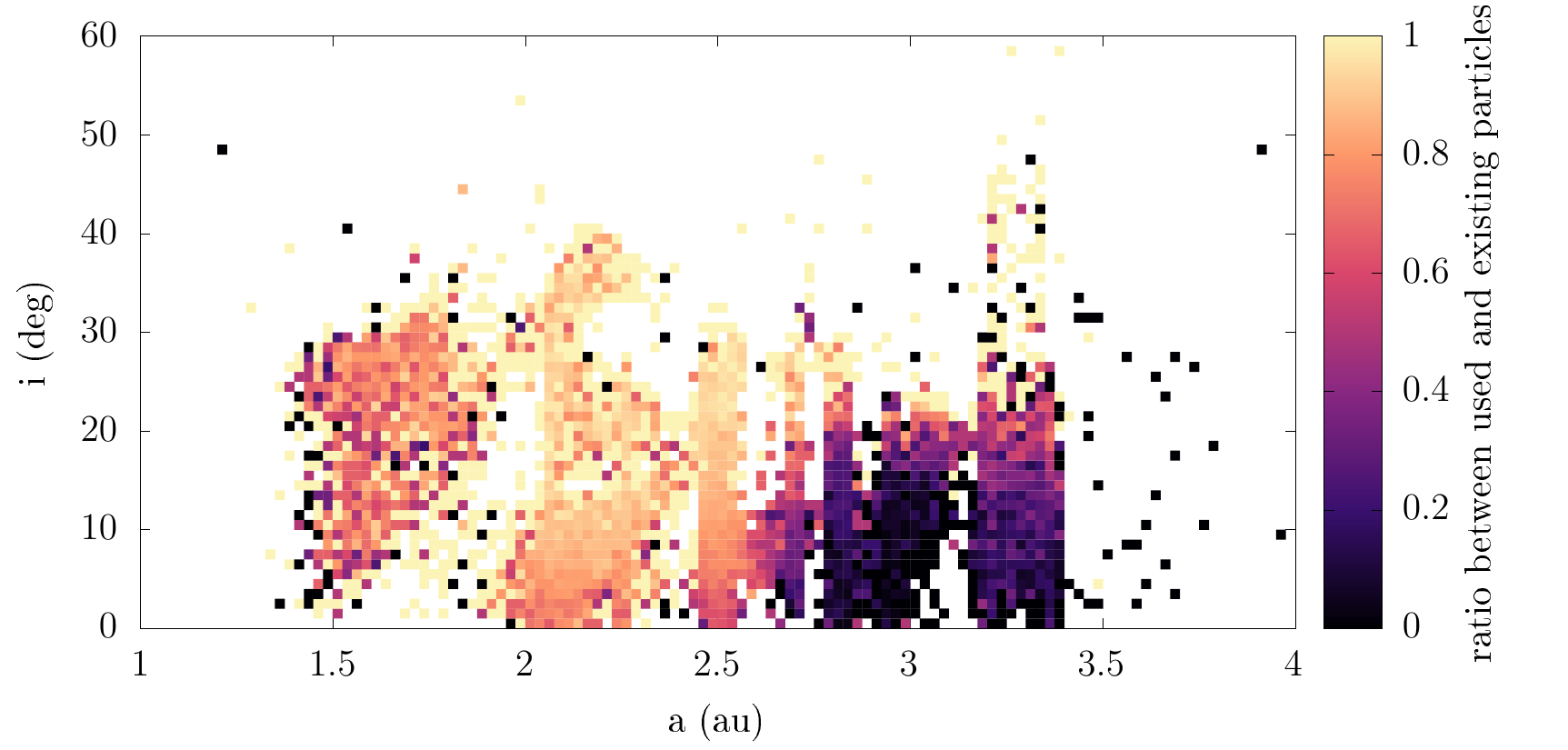}
\caption{A heat map showing how efficiently test asteroids are delivered to orbits that have close encounters with the Sun as a function of initial $a$ and $i$. The color coding ranges from black (when no asteroids from a particular cell are used) to yellow (when all test asteroids reach $q^*$). }
\label{fig:heatmap}
\end{figure}

For this analysis, we started from the 80,667 test asteroids that have previously been integrated during the development of the models by \citet{Granvik2016} and \citet{Granvik2018}. We found that the $q$ of 55,253 of these test asteroids ($\sim70$ per cent) eventually reached below $q^*$, and we therefore focus our analysis on this subsample. The other test asteroids collided with a planet, were ejected from the system after a close encounter before reaching $q^*$, or their integration was stopped after returning to the main asteroid belt and remaining there for more than a few Gyr. In Fig.~\ref{fig:heatmap}, we show a heat map of the ratio between the asteroids used and the entire simulated population. As expected, asteroids originating in the outermost part are delivered less efficiently close to the Sun \citep[see, e.g.,][]{Granvik2018}.  

The resonance-detecting algorithm that we developed is able to identify occurrences of resonances throughout the entire time evolution of the test asteroids. Here, we will only consider the last mechanism that drove each test asteroid below $q^*$. In total, 21,347 NEAs were driven close to the Sun due to an MMR ($\sim39$ per cent) and 29,780 due to a secular resonance ranging from $\nu_2$ to $\nu_6$ ($\sim54$ per cent). Finally, for 4,126 out of 55,253 particles ($\sim7$ per cent), none of the resonant mechanism that we consider has been identified as responsible for decreasing their $q$ However, for 312 of them, $\omega$ was found to be librating around either $90\deg$, or $270\deg$. 

\begin{figure*}
\centering
\includegraphics[width=\textwidth]{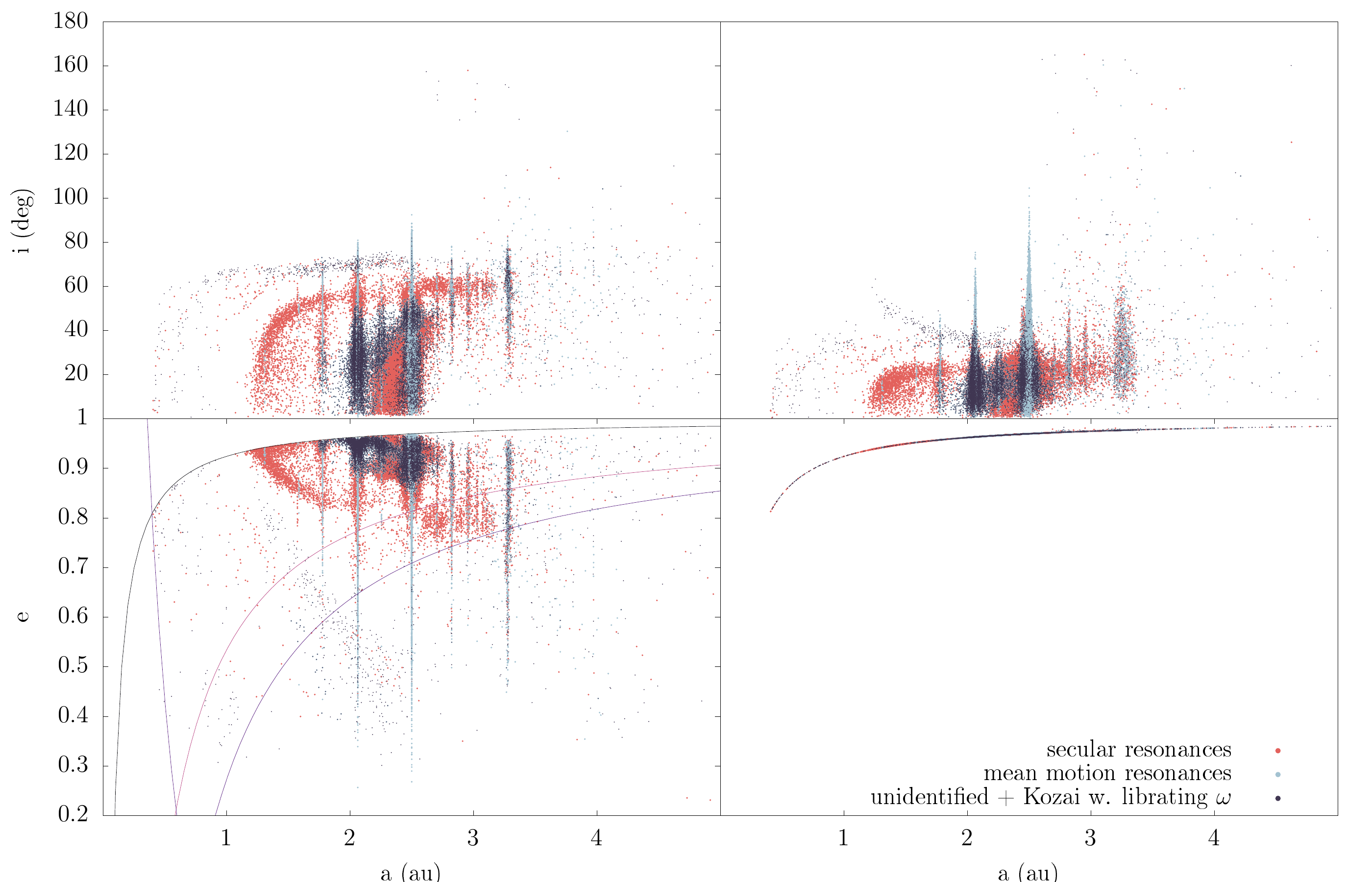}
\caption{Top panels: The $a-i$ distribution of the averaged orbital elements (left) and osculating orbital elements (right) for all test asteroids that have reached below $q^*=0.076\au$. The colour coding corresponds to the resonance responsible for reducing $q$ below the critical value: blue for MMRs and red for secular resonances. The asteroids that have not been flagged with a mechanism, as well as asteroids in the Kozai resonance with librating $\omega$ are plotted with dark blue. Bottom panels: The distribution of test asteroids in the $a-e$ plane according to the last recorded mechanism. The left panel shows the averaged $e$ while the right shows the osculating $e$.  In addition, in the left panels we show the location of the perihelion and aphelion distances of Venus (light purple and magenta), the perihelion distance of Mars (dark purple), and the location of $q=q^*$ (black). These parameters help shape the final orbital distribution of the test asteroids.  }
\label{fig:all}
\end{figure*}

In Fig.~\ref{fig:all}, we show the distribution of the test asteroids that eventually reached below $q_*$ in the $(a,e)$ and $(a,i)$ planes. We focus on the range $0<a<5\au$ since there are very few particles with larger $a$. The right panels correspond to the osculating orbital elements when $q=q^*$ --- computed by performing a linear interpolation between two time-steps: the last time-step with $q>q^*$ and the first with $q<q^*$. The left panels show 'averaged' orbital elements. We considered  the last two periods of the oscillations of $e$, just before $q=q^*$. This was determined by finding the two last minimum values in the time evolution of $e$, which occurs when the slope of $e(t)$ changes sign, after acquiring a minimum absolute value. We then took the middle point between the two extreme values that were recorded (the maximum and the minimum). Note that this method is not foolproof, especially if there are close encounters during the considered time window. However, for our purposes it is enough to obtain an overview of the `average' values of the orbital elements.

\begin{figure}
\includegraphics[width=0.5\textwidth]{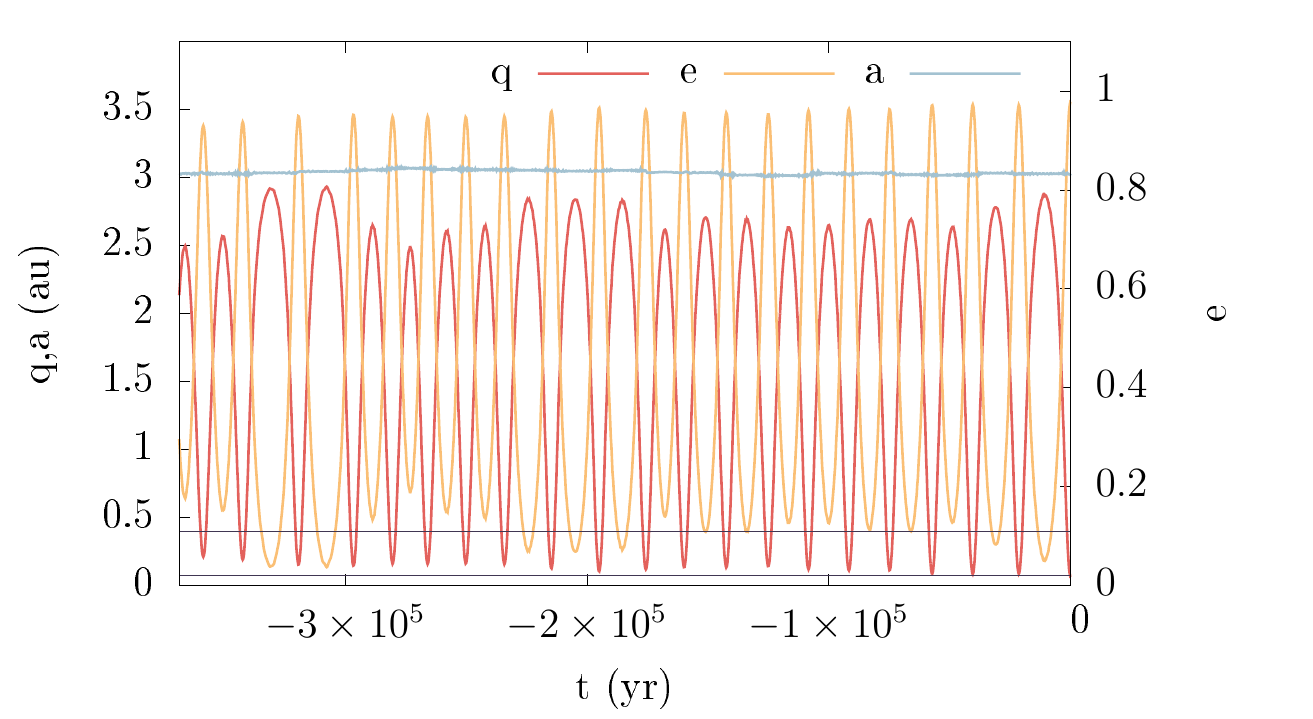}
\caption{ An example of the final stages of the time evolution of $q$ (red), $e$ (yellow) and $a$ (blue), of a test asteroid that has large oscillations in eccentricity, taking it in and out of the near-Earth region within the same period. At $t=0$ $q$ crosses $q^*$. The horizontal lines corresponds to $q_l$ and $q^*$.}
\label{fig:large_osc}
\end{figure}

An interesting feature in 'average' $e-a$ plot is that there is a clear cut-off at $e=0.5$, which means that for $a>2.6\au$ there are test asteroids that have an `average' $e$ that places them outside the near-Earth region. This is a result of large oscillations in the $e$ of certain test asteroids (between $\sim0$ and $\sim1$ in the most extreme cases), that gives an `average' $e$ of 0.5. Consequently, it is evident that not all test asteroids are NEAs at all times. In Fig.\ref{fig:large_osc} we show a test asteroid that exhibits such a dynamical behaviour.

\begin{figure}
\centering
\includegraphics[width=0.48\textwidth]{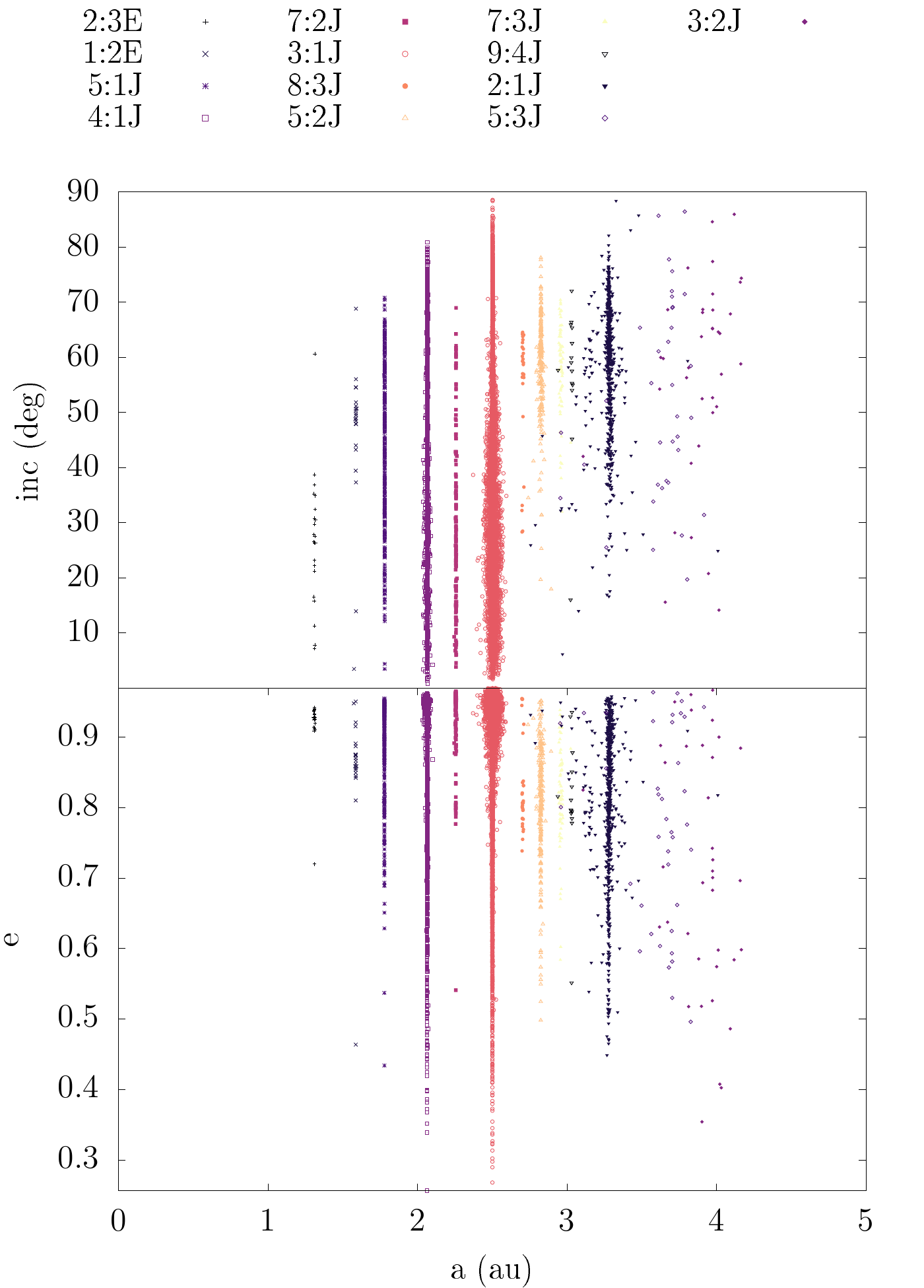}
\caption{The $a-i$ (top) and $a-e$ (bottom) distribution of test asteroids that have, due to capture into an MMR, evolved to orbits that have close encounters with the Sun. Each MMR is plotted with a different combination of colours and point types. The difference between the location of an MMR and the averaged $a$ for some cases is the result of a close encounter in the final stages of the asteroid's orbital evolution. Here we do not make a distinction between cases of MMRs overlapping with secular resonances.}
\label{fig:mmr}
\end{figure}

Next, we show the location for each MMR considered in our study. In Fig.~\ref{fig:mmr}, we plot the distribution of the averaged orbital elements (computed with the method described above) in the $(a,i)$ plane of test asteroids that have been flagged as being in an MMR at their end state. The structure and location of each MMR is very well defined, but there are noticeable points where there is a difference between the theoretically predicted location of the MMR and the calculated averaged orbital elements. In the case where a close encounter occurs at the latest stages of the evolution of a test asteroid, its $a$ suffers a kick, which moves it out and away from the MMR. The resonant argument however may not have had enough time to make a full circulation and thus prevents our tool from identifying the NEA as being outside of the resonance. We found that $14717$ test asteroids are in the 3:1J MMR during their last stages (27 per cent) and $5067$ (9 per cent) are in the 4:1J MMR. The remaining test asteroids are divided among the other MMRs. 

\begin{figure}
\centering
\includegraphics[width=0.48\textwidth]{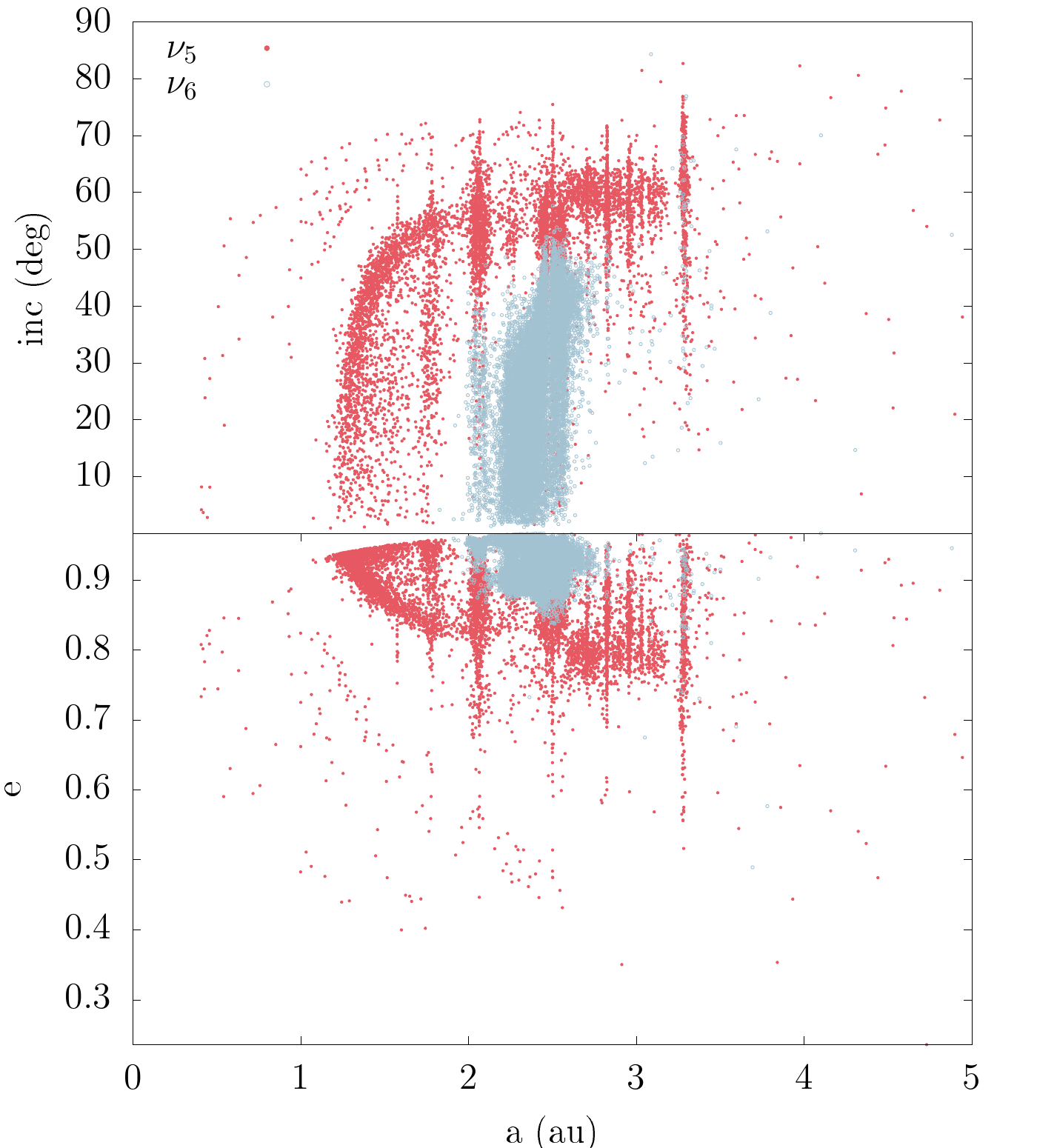}
\caption{The 'average' $a-i$ (top panel) and $a-e$ (bottom panel) distributions of test asteroids for which either the $\nu_5$ (red) or $\nu_6$ (blue) secular resonance has effectively driven their $q<q^*$.}
\label{fig:sr}
\end{figure}

Secular resonances are harder to identify than MMRs, because the `shape' corresponding to the libration of the resonant argument can be more complex. To this end, we use a wide range of window sizes. In Fig.~\ref{fig:sr}, we show that the orbital distributions of test asteroids marked as being in the $\nu_5$ or $\nu_6$ secular resonances at the end of their evolution have a more diffuse layout. The existing analytical maps of the locations of these resonances are of no use here, since these NEAs have large $e$ and such maps cannot be derived to compare with our results. $15044$ test asteroids have been flagged as being in the $\nu_6$ secular resonance when $q=q^*$, while $7679$ have been flagged as being in $\nu_5$. There are also $570$ test asteroids that have been flagged as being in both the $\nu_5$ and the $\nu_6$ secular resonances. These test asteroids can, of course, not simultaneously be in both resonances. We have ensured that such an artefact is indeed avoided in our analysis by calculating their $g$ frequency and comparing it with $g_5$ or $g_6$ (section~\ref{sec:methods}). However, the use of different-sized windows implies that our method identifies both these resonances in cases when these test asteroids may, at different times, have been in either $\nu_5$ or $\nu_6$. We split these test asteroids equally among the $\nu_5$ and $\nu_6$ groups, increasing the number of test asteroids in these groups to 15329 ($\sim28$ per cent) and 7964 ($\sim14$ per cent), respectively. 

\begin{figure}
\centering
\includegraphics[width=0.48\textwidth]{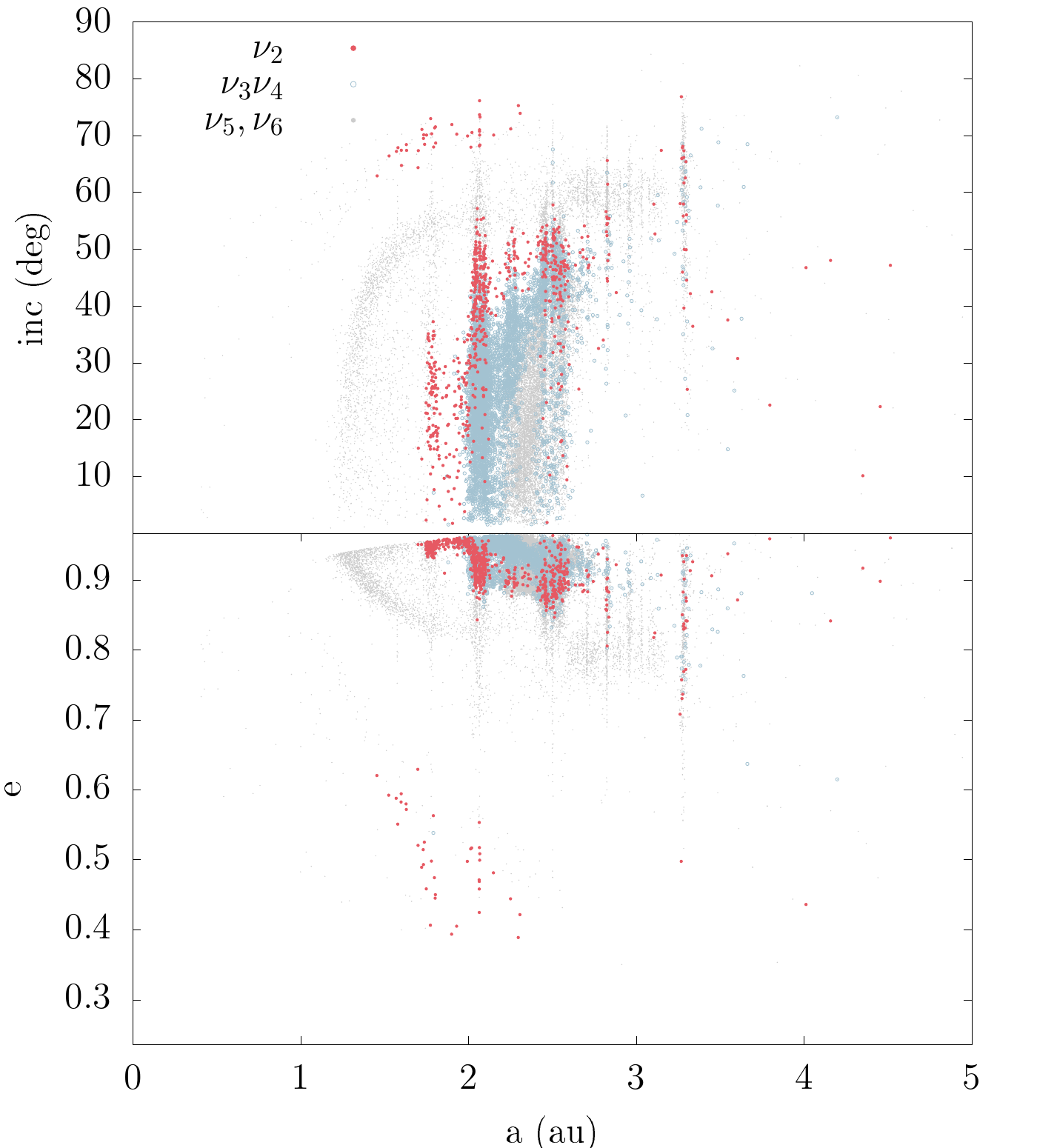}
\caption{The 'average' $a-i$ (top panel) and $a-e$ (bottom panel) distribution of asteroids that either the $\nu_2$ (red) or $\nu_3/\nu_4$ (blue) secular resonance has effectively driven their $q<q^*_\text{dis}$. For comparison, we also plot the location of the $\nu_5$ and $\nu_6$ secular resonances in grey.}
\label{fig:sr_terr}
\end{figure}

Secular resonances with the $2^\text{nd}$, $3^\text{rd}$ and $4^\text{th}$ apsidal eigenfrequencies of the Solar System, corresponding to the terrestrial planets, Venus, Earth and Mars, are shown in Fig.~\ref{fig:sr_terr}. Specifically, $5194$ test asteroids have been identified as being in the $\nu_3$ or $\nu_4$ secular resonances, while $728$ as being in the $\nu_2$ secular resonance. $565$ asteroids were marked as having been in all three resonances during the latest stages of their evolution. We split those test asteroids among the two groups, similar to what we did in the case of the $\nu_5$ and $\nu_6$ secular resonances, leading to $5477$ asteroids ($\sim10$ per cent) in the $\nu_3\nu_4$ resonance and $1010$ ($\sim2$ per cent) in the $\nu_2$ resonance. Since we allow a range of $g_\text{pl}\pm1.5\arcsecyr$ in the calculated $g$ frequency of the test asteroids for secular resonances of the terrestrial planets, and $g_\text{pl}\pm3\arcsecyr$ for secular resonances with Jupiter and Saturn, there is an overlap between $\nu_2$ and $\nu_5$, with $g_2\simeq7.5\arcsecyr$ and $g_5\simeq4.3\arcsecyr$, respectively. This is obvious in Fig.~\ref{fig:sr_terr} where a few high-$i$ NEAs in $\nu_2$ possibly belong to $\nu_5$.   

Let us now look at the 4126 test asteroids that were not identified as being in a resonance at the moment when they reached the end of their evolution. We found that, in many cases, some resonance was able to raise the $e$ of a test asteroid significantly and thereafter the secular oscillations induced by the Kozai mechanism were enough to allow it to pass the $q^*$ threshold. This can also be deduced from Fig.~\ref{fig:all}. These plots show that some of the dark blue test asteroids had been trapped in an MMR, but during the last time-steps of the integration they have escaped. In addition, there are test asteroids that are pushed towards the Sun after a close encounter with a planet. There can also be other explanations such as higher order mean-motion and secular resonances that we do not take into account in this study. Finally, it could be that our algorithm might fall short of identifying a libration, because the accuracy of our method is less than 100 per cent. In any case, we categorise all these test asteroids as 'unidentified+Kozai' as they comprise all the test asteroids for which no resonant mechanism has been identified and the ones for which $\omega$ was found to be librating. The reason we do not separate test asteroids with librating $\omega$ and test asteroids with unidentified mechanisms in two separate categories, is that even if $\omega$ is not librating but circulating, large enough variations in $e$ and $i$ can occur. An overview of the most important identified resonances is given in Table~\ref{tab:mechanisms}.
 
There is an apparent dearth of test asteroids with small $i$ ($i<25^{\circ}$) and $a\geq 2.65$~AU (Fig.~\ref{fig:all}). As the $e$ of test asteroids in that region increases, their aphelion distance increases as well and, as a result, they reach or cross the orbit of Jupiter, suffering close encounters that typically eject them from the inner Solar System. However, asteroids with large $i$ can be more long-lived despite having a large aphelion distance, because the likelihood for close encounters is lower as a result of spending more time far from the ecliptic plane, and because they also have higher encounter velocities. The large variation in the average $e$ values can be explained by a very fast increase in the osculating $e$ of test asteroids, especially those trapped in an MMR, or by large oscillations in the osculating $e$ from the Kozai mechanism. 

\begin{table}
\centering
\caption{The number of test asteroids in the resonance that brought the $q$ of the test asteroids below $q_l$, $N_{q_l}$ (second column), the predominant resonances during $\tau_{lq}$, $N_\text{evo}$ (third column), and the last identified resonance that reduces $q$, $N_{q^*}$ (fourth column). The last column is the number of asteroids for which the last identified resonance was the same as the predominant one during their evolution in near-Sun space, $N_\text{sm}$ .}
\begin{tabular}{ccccc}
\hline
mechanism & $N_{q_l}$ & $N_\text{evo}$ &  $N_{q^*}$ & $N_\text{sm}$  \\
\hline
$\nu_6$      & 8885  & 11389 & 15329  & 9902  \\
3:1J MMR     & 18512 & 16761 & 14717  & 11673 \\
$\nu_5$      & 3708  & 5371  & 7964   & 3811  \\
$\nu_3\nu_4$ & 6115  & 5070  & 5477   & 2915  \\
4:1J MMR     & 5670  & 3572  & 5067   & 2149  \\
unidentified+Kozai          & 9729  & 10882 & 4126   & -     \\
$\nu_2$      & 673   & 385   & 1010   & 71    \\
2:1J MMR     & 1183  & 1181  & 670    & 546   \\
other        & 778   & 642   & 893    & 160   \\ 
\hline
total & 55253 & & & 31227     \\
\end{tabular}
\label{tab:mechanisms}
\end{table}

\subsection{Orbital evolution of NEAs with small $q$: time-scales and predominant dynamical mechanisms }

Let us first consider the effective lifetime of NEAs in orbits with small $q$, which is defined here as $q<q_l=0.4\au$. The recorded lifetime, $\tau_{lq}$, ranges from a few hundred years to $\sim6 \times 10^8$~yr. For 27 particles, $q$ was already lower than $q_l$ when they entered the near-Earth region from the asteroid belt, so for those $\tau_{lq}$ is calculated from that moment on.

We combine information about resonance occurrences coming from windows of every size and, thus, have records for the orbital history of every asteroid at each point of its evolution in the near-Earth region. Focusing on the time interval starting when $q<q_l$ for the first time and continuing until $q \le q^*$ (the $\tau_{lq}$) we determine which resonance (if any at all) the test asteroid was trapped in for most of the time. That is then considered to be the predominant resonance during $\tau_{lq}$. In Table~\ref{tab:mechanisms} we present the predominant resonance during $\tau_{lq}$, as well as the resonance that was active at $q=q_l$. We then compare the predominant mechanism and the last recorded mechanism presented in the previous subsection (Sec.~\ref{sec:last}) by counting the number of cases in which they have been the same. From that comparison, we exclude asteroids for which no resonance was identified, because the underlying mechanism is uncertain.

\begin{figure}
\centering
\includegraphics[width=0.48\textwidth]{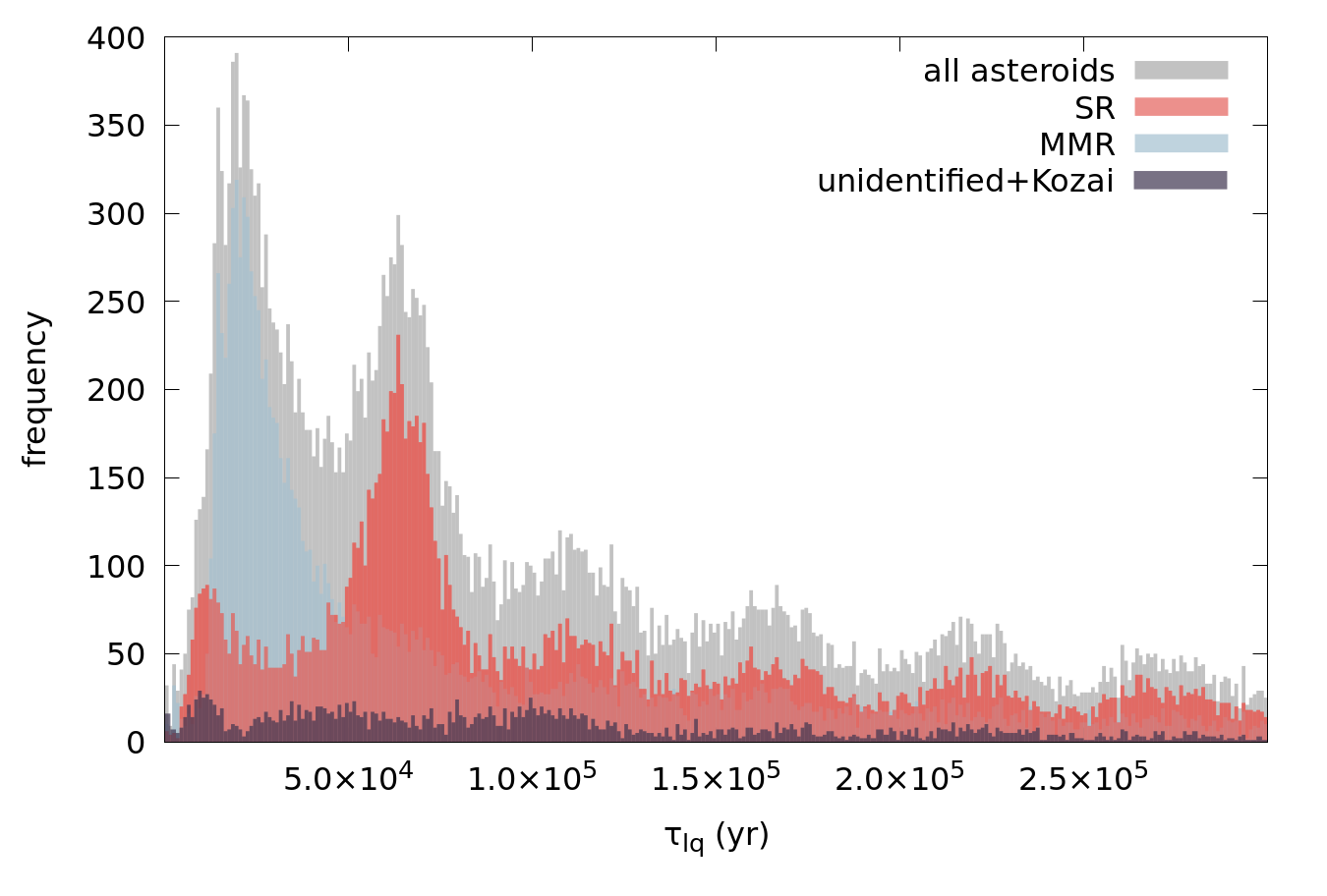}
\caption{Histogram of the recorded time span from the first time $q<q_l$ until it reaches $q=q^*$ ($\tau_{lq}$) for all test asteroids (grey) and also the respective histograms for asteroids that have been flagged as being in a secular resonance (blue), an MMR (red) and the histogram for test asteroids which are in the Kozai resonance, as well as those for which the $q$-reducing dynamical mechanism is unidentified (dark blue). Each bin has a width of $1000\yr$.}
\label{fig:hist_all}
\end{figure}

\begin{figure}
\centering
\includegraphics[width=0.48\textwidth]{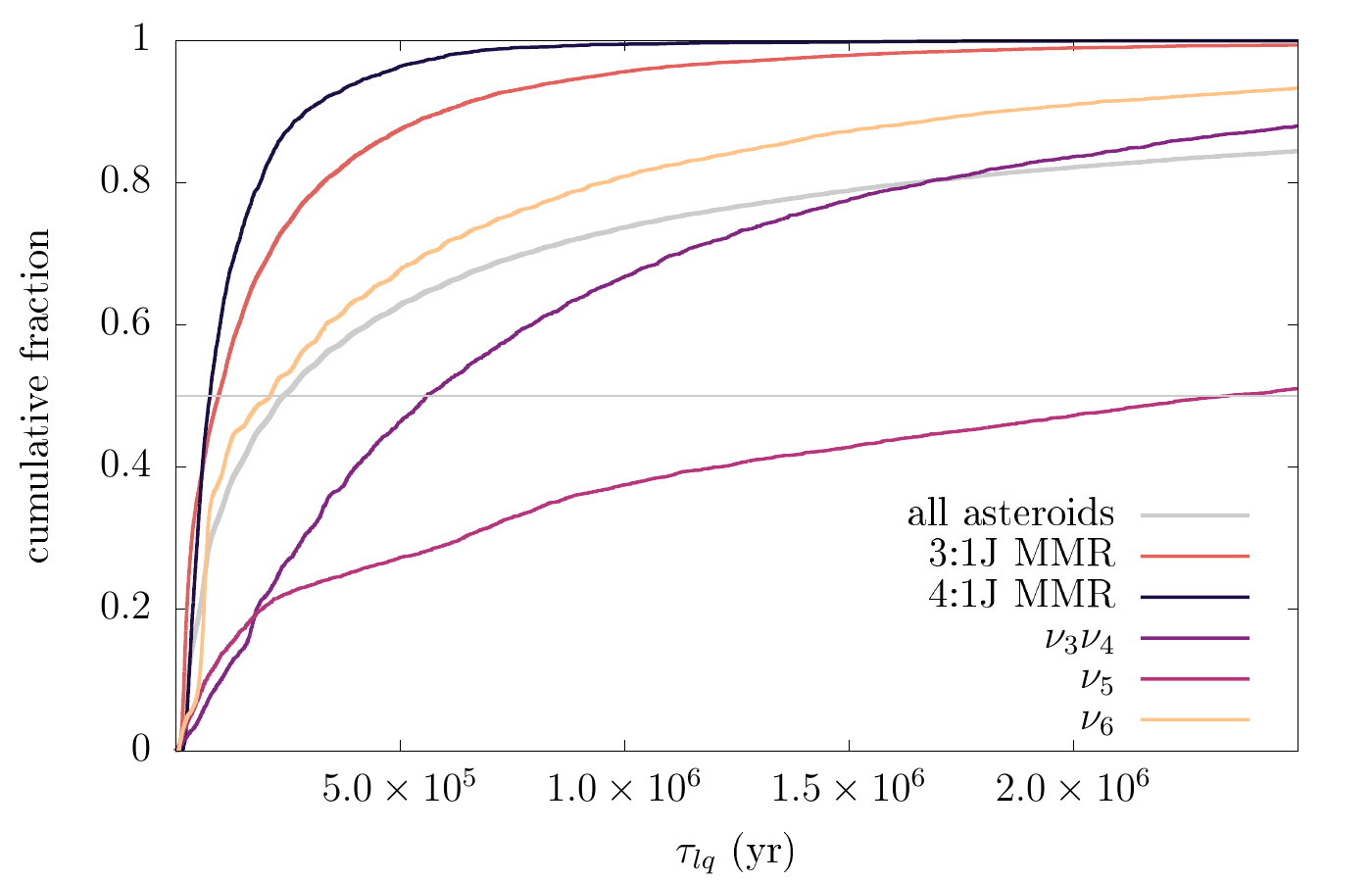}
\caption{The cumulative distributions of the $\tau_{lq}$ time-scales for the most important resonances that decrease $q$: 3:1J (red) and 4:1J (blue) MMRs and the $\nu_3\nu_4$ (purple), $\nu_5$ (magenta) and $\nu_6$ (orange) secular resonances. We also plot the cumulative distribution for all test asteroids (gray).}
\label{fig:cumul_pop}
\end{figure}

In Fig.~\ref{fig:hist_all}, we show a histogram of the $\tau_{lq}$ of all test asteroids covering the range from 0 to $3 \times 10^5\yr$, as more than 50 per cent of the $\tau_{lq}$ time-scales fall in this range. There are two distinct peaks in the histogram corresponding to test asteroids that had either an MMR or a secular resonance as the predominant mechanism during $\tau_{lq}$. As we can see from the same plot, MMRs are responsible for the first, strongest peak, and secular resonances are responsible for the second peak. This translates to MMRs being a faster mechanism for bringing NEAs closer to the Sun. In Fig.~\ref{fig:cumul_pop}, we show the cumulative distribution of $\tau_{lq}$ time-scales for the most important mechanisms, namely the 3:1J and 4:1J MMRs and the $\nu_3\nu_4$, $\nu_5$ and $\nu_6$ secular resonances. We note that the 4:1J MMR is the fastest mechanism, while $\nu_5$ is the slowest.

There is also an apparent periodicity of $\sim5\times 10^4$~yr in the subsequent, weaker peaks in the histogram, especially coming from particles in a secular resonance. This is related to the $g_6\simeq28.6\arcsecyr$ secular frequency of the Solar System which, in the set of simulations that we are using, includes gravitational perturbations by all planets as well as Pluto and the Moon. In particular, considering that the time evolution of $q$ can be roughly described by $q(t)=A \sin(g_6t+\phi)-(1/\tau) t$, where $A$ is the amplitude of the oscillation, $\phi$ the phase and $\tau$ the time-scale of the decay, then $q$ crosses $q_l$ when the sine function is on its "decreasing" branch, that is, when the angle is between $\pi/2$ and $3\pi/2$. Similarly, it crosses $q^*$ on a similar phase. This results in peaks in the $\tau_{lq}$ histogram that are multiples of the period corresponding to the $g_6$ frequency.    

\subsection{Average time spent at different $q$ and $r$}
\label{sec:rel_qr}

Following the method described in Sec.~\ref{sec:methods}, we can compute the total time that a test asteroid spends in various $q$ intervals from the first moment its $q<q_l$. Note that the sum of all these time-spans does not always add up to $\tau_{lq}$, since $q$ can rise above $q_l$ during the evolution of an asteroid. Next, we divide the calculated time with the width of each bin and compute the time density, $T_q$. Finally, we split the test asteroids in groups according to the predominant resonance during $\tau_{lq}$, and compute the average time densities over all the test asteroids identified with the same predominant resonance, $\widetilde{T_q}$. 

\begin{figure*}
\centering
\includegraphics[width=\textwidth]{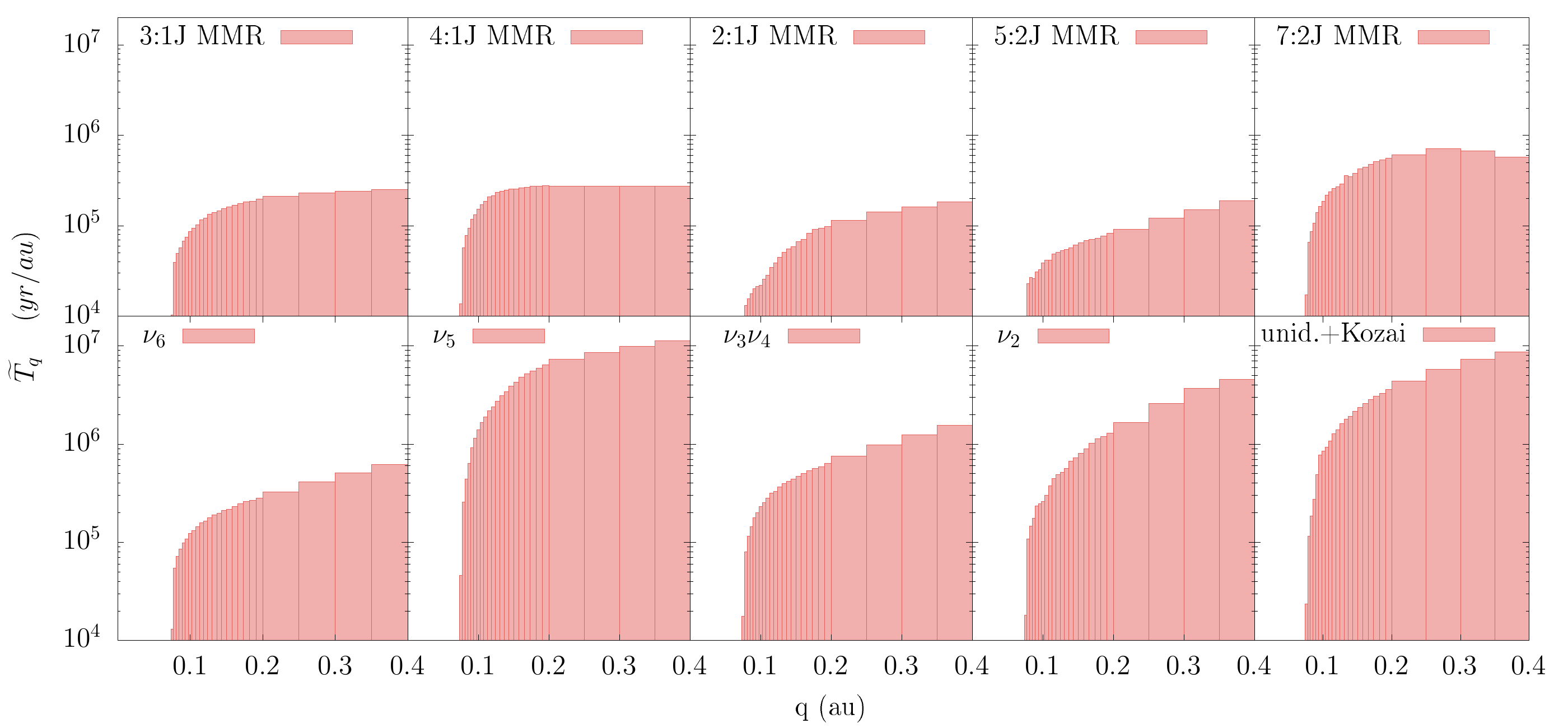}

\caption{The distribution of $\widetilde{T_q}$, that is, the time that asteroids spend on orbits with different $q$, averaged over test asteroids and normalized by bin width, for the cases in which the predominant resonance during the $\tau_{lq}$ evolution of the test asteroids was one of the 3:1J, 4:1J, 2:1J, 5:2J or 7:2J MMRs, the $\nu_2$, $\nu_3\nu_4$, $\nu_5$ or $\nu_6$ secular resonances, and finally, test asteroids that are in the Kozai resonance or for which the $q$-reducing mechanism is unidentified}. 
\label{fig:hist_binsq_dis}
\end{figure*}

The result is shown in Fig.~\ref{fig:hist_binsq_dis} for the most important resonances during $\tau_{lq}$, that is, 3:1J, 4:1J, 2:1J, 5:2J, and 7:2J MMRs as well as the $\nu_2$, $\nu_3\nu_4$, $\nu_5$, and $\nu_6$ secular resonances. We also include asteroids for which no predominant mechanism was recorded. A larger $\widetilde{T_q}$ value tells us that the resonance drives the $q$ of an asteroid close to the Sun more slowly. However, the total time spent in each of these bins also depends on the location of the resonance. For example, the $e$ of asteroids trapped in 2:1J rises slower than of those trapped in 4:1J but asteroids in 2:1J spend less time at small $q$.

\begin{figure*}
\centering

\includegraphics[width=\textwidth]{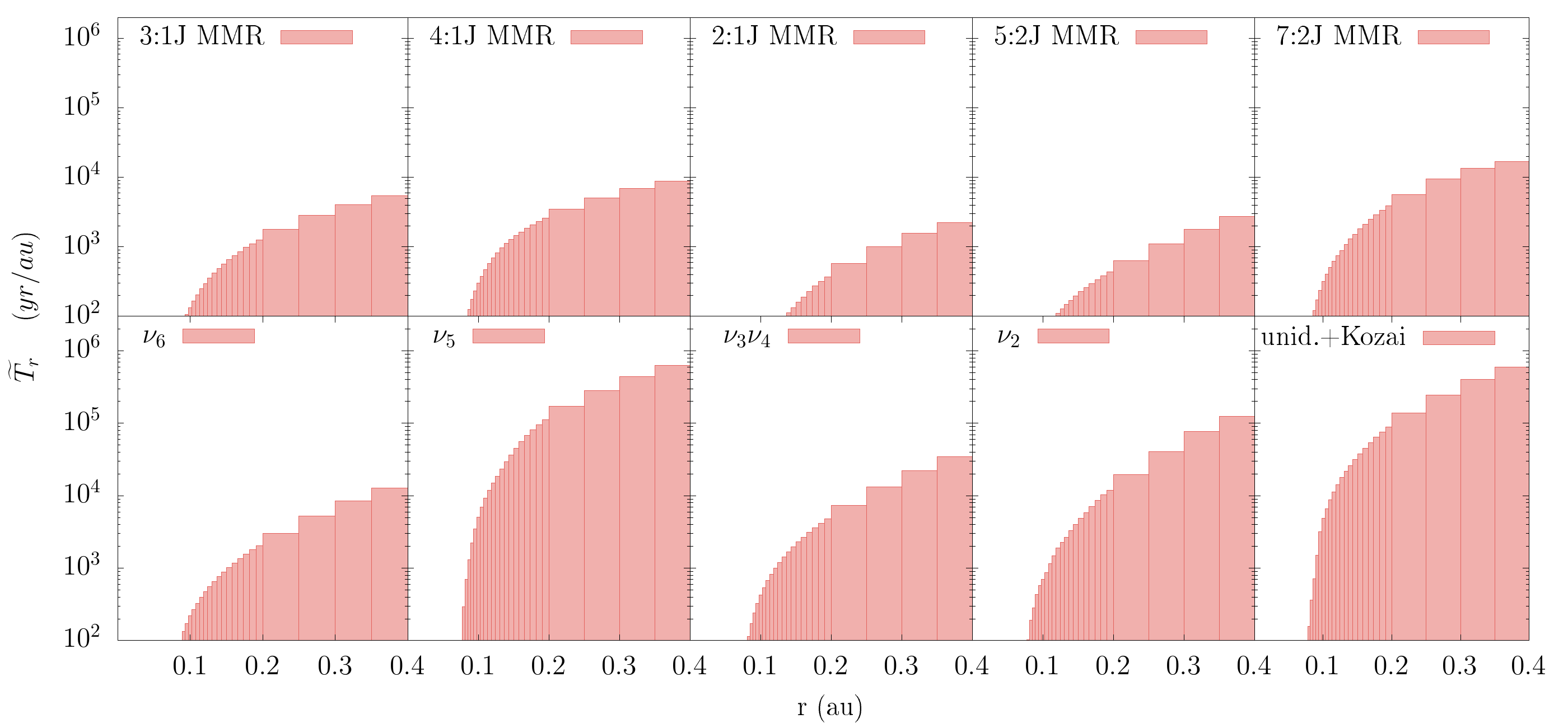}
\caption{The same as Fig.~\ref{fig:hist_binsq_dis}, but for the time spent at different heliocentric distances averaged over all test asteroids that are predominantly associated with the most important resonances and divided by bin width, $\widetilde{T_r}$.}
\label{fig:hist_binsr_dis}
\end{figure*}

Similarly to $\widetilde{T_q}$, we can calculate how much time, on average, an NEA spends in a certain heliocentric distance interval during its dynamical evolution after the first time $q<q_l$, $\widetilde{T_r}$ (Fig.~\ref{fig:hist_binsr_dis}).

\subsection{Efficiency of resonant mechanisms in decreasing $q$} 

One of the aims of the present study is to get an estimate on the efficiency of each resonance in reducing $q$, using the extensive data produced by the aforementioned numerical simulations. This can be done by determining the amount of time asteroids tend to spend trapped in each resonance during their evolution and measure the change in their $q$ during that time. Thus, we define the rate of change of $q$, $\frac{\Delta q}{\Delta t}$ as a metric of the efficiency of a resonance. We treat overlapping resonances by distinguishing between whether MMRs occur together with $\nu_5$ and $\nu_6$, or not, and prioritising $\nu_5$ and $\nu_6$ over $\nu_2$, $\nu_3$ and $\nu_4$. 

To measure the time a test asteroid is trapped in a resonance, we locate the start and end of a resonance using a so-called `master table', in which we project the time evolution of the resonances identified by analyzing all window sizes. Each entry of the master table corresponds to the half the smallest window size (15 output time-steps). In this way, a resonance identified in a window of size e.g. 100 output time-steps will spread over seven entries of the master table. We find the dwell time in a particular resonance by multiplying the number of consecutive entries listing the same resonance by 15---the number of output time-steps in each entry---and by $250\yr$---the output time-step.

Measuring the change in $q$ is a more complex task. The forced frequency imposed on an asteroid, which manifests in the oscillation of $q$, is different for each resonance, complicating the choice of a single recipe such as averaging $q$ using a moving window of constant size. We address this issue in the following way. For each entry in the master table of the resonances identified for an asteroid, we focus on the time evolution of $q$ for the same asteroid. We locate the closest local maximum and minimum, starting from the middle of each time-span, i.e., the center of the time interval of each entry. We then take the middle value between the acquired local extrema values. This serves as the 'average` value of $q$. We measure the efficiency by taking the difference between the 'average` value, when the test asteroid enters and exits a resonance, and divide it with the total time spent in the resonance. 

\begin{figure}
\centering
\includegraphics[width=0.48\textwidth]{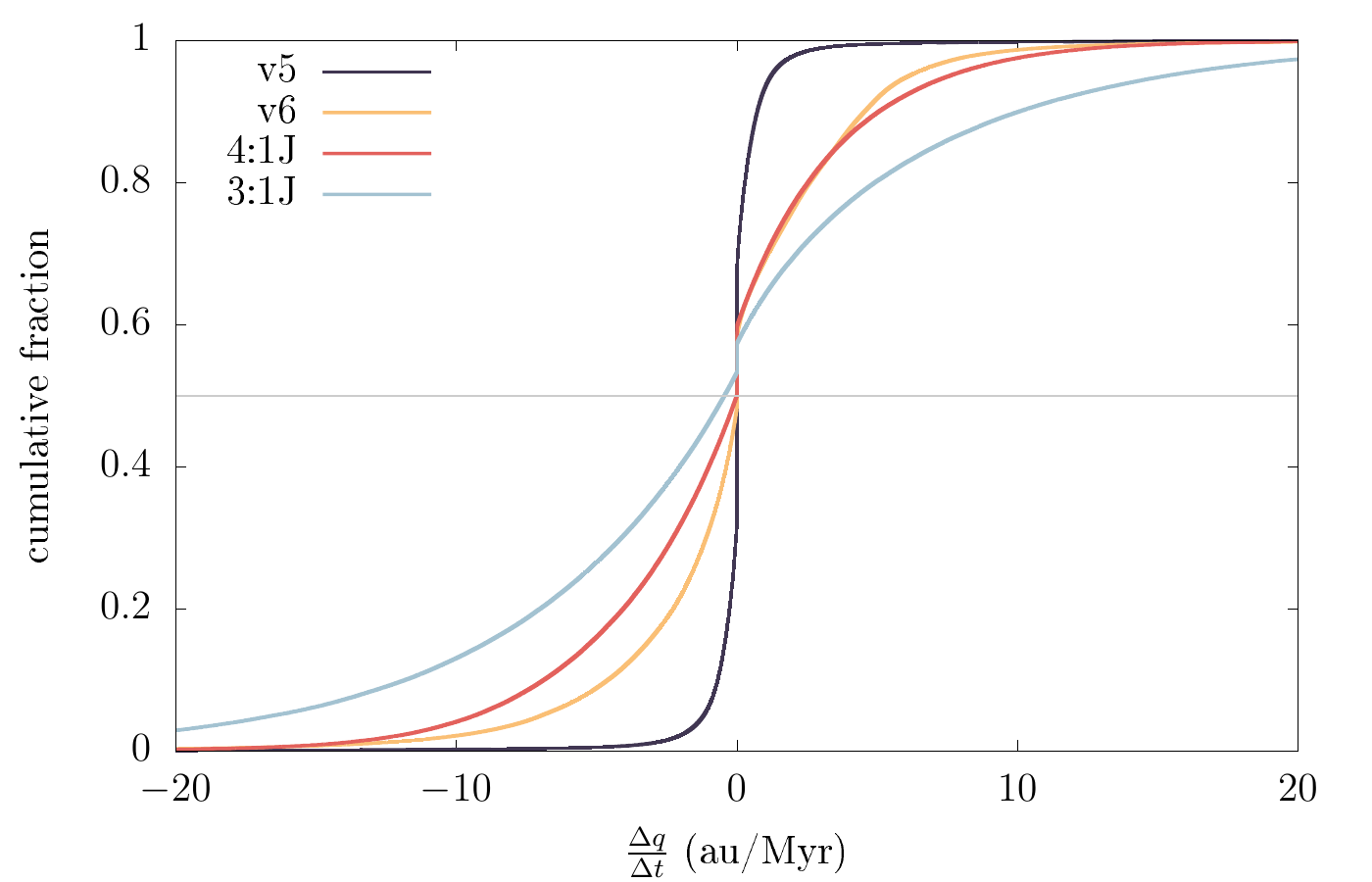}
\caption{The cumulative distributions of the rate of change of $q$, $\frac{\Delta q}{\Delta t}$, for the most important mechanisms reducing $q$: the $\nu_5$ (dark blue) and $\nu_6$ (yellow) secular resonances, and the 3:1J (light blue) and 4:1J (red) MMRs. The gray line corresponds to the median of the calculated $\frac{\Delta q}{\Delta t}$ for each resonance.}
\label{fig:efficiency}
\end{figure}

We keep record of every occurrence of each resonance for all test asteroids and their efficiencies. In Fig.~\ref{fig:efficiency}, we show $\frac{\Delta q}{\Delta t}$ for the four most important resonances. We find that 3:1J is the most efficient mechanism, followed by 4:1J, $\nu_6$, and, finally, $\nu_5$. In addition, $\sim57$ per cent of the test asteroids in 3:1J have $\frac{\Delta q}{\Delta t}\le0$ (and $\sim53$ per cent have $\frac{\Delta q}{\Delta t}<0$). Similarly, for 4:1J this number is $\sim60$ ($\sim50$) per cent, for $\nu_6$ $\sim58$ ($\sim48$) per cent and for $\nu_5$ $\sim68$ ($\sim32$) per cent. Note, however, that the test asteroids that we study only include the ones that have reached very close to the Sun so their preferential evolution was that of increasing their $e$.

The explanation for the relatively large difference between two percentages calculated for the same resonance is that we have not imposed a minimum number of consecutive entries during which a test asteroid has to be identified as being in a resonance, with the minimum value in the current method being 2, which translates to $7.5\times 10^3\yr$. This, combined with the fact that the perihelion distances are given with only 4 decimal places, implies that in a significant fraction of the cases $\Delta q$ is too small to be measured and hence $\frac{\Delta q}{\Delta t}=0$. Applying a filter on the minimum number of consecutive windows will quantitatively change the cumulative distribution but will not qualitatively affect the curves of the resonances when compared to each other. Consequently, the results obtained without imposing a threshold value serve well as a proxy when comparing different mechanisms, although the computed rates may not be accurate. Finally, there is a chance that the results are skewed towards larger rates, because we have not kept track of any close encounters happening while a test asteroid is in a resonance, which may also change $a$ and hence $q$. 

\begin{figure}
\centering
\includegraphics[width=0.48\textwidth]{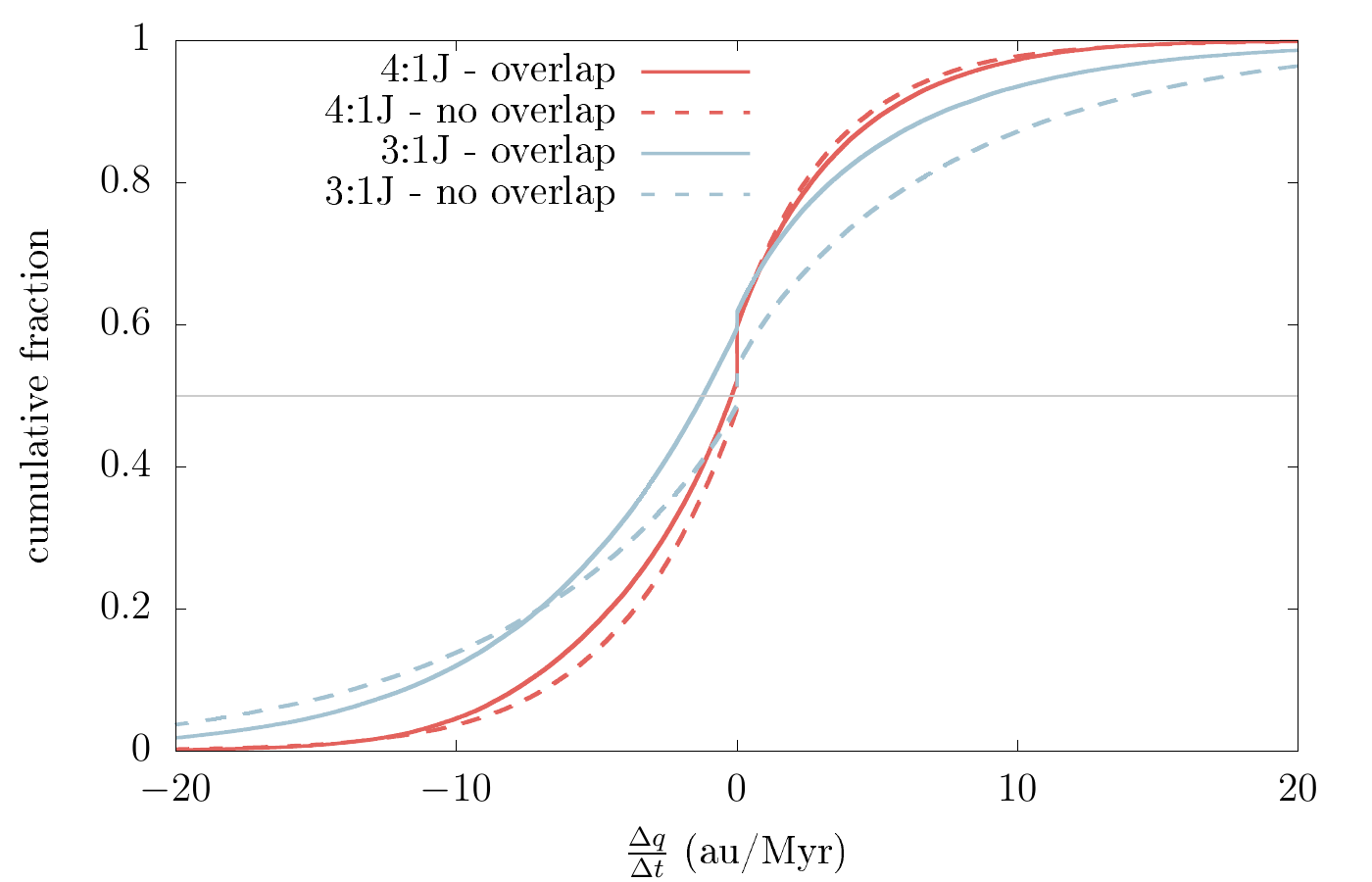}
\caption{The cumulative distributions of $\frac{\Delta q}{\Delta t}$ for the 4:1J (red) and 3:1J (blue) MMRs  $q$ when overlapping with other resonances (solid line) or when there is no overlap (dashed line). The gray line corresponds to the median of the calculated $\frac{\Delta q}{\Delta t}$ for each resonance.}
\label{fig:efficiency_overlap}
\end{figure}

In Fig.~\ref{fig:efficiency_overlap}, we show the rate of change of $q$ for the 3:1J and 4:1J MMRs, making a distinction between occurrences when only each MMR was in effect and occurrences of other resonances overlapping with these MMRs. We find that, for the 3:1J MMR, the presence of a secular resonance has a more pronounced effect in reducing $\frac{\Delta q}{\Delta t}$. 

\subsection{ER-specific mechanisms and time-scales for reducing $q$}
\label{sec:ERS}

\citet{Granvik2018} define six distinct escape routes (ERs) that NEAs follow from the asteroid belt into the near-Earth region: (i) the 2:1J complex, which consists of the 2:1J, 9:4J and 11:5J MMRs as well as the $2(g-g_6) + s-s_6$ secular resonance, commonly referred to as z$_2$ , (ii) the 3:1J complex, consisting of the 3:1J MMR and the outer $\nu_6$ secular resonance, (iii) the 5:2J complex that includes in addition the 7:3J and 8:3J MMRs, (iv) the $\nu_6$ complex comprised of the inner $\nu_6$ secular resonance and the 4:1J and 7:2J MMRs, (v) the Hungarias, and (vi) the Phocaeas. Note that the latter two are akin to source regions, and the escape routes are typically the neighboring resonances.  \citet{Granvik2018} provide information for the ER taken by each test asteroid in the set of simulations. By taking into account only the largest asteroids with absolute magnitude $H<14.4$~mag, we get a total number 44,731 of test asteroids, the $q$ of which eventually evolved below $q^*$. Out of the 44,731 test asteroids, only 1,949 ($\sim4$ per cent) entered the near-Earth region through the 2:1J complex, 8,951 ($\sim20$ per cent) through the 3:1J complex, 2,952 ($\sim7$ per cent) through the 5:2J complex, 16,482 ($\sim37$ per cent) through the $\nu_6$ complex, 6,644 ($\sim15$ per cent) used to belong in the Hungarias and 7,753 ($\sim17$ per cent) in the Phocaeas.

An overview of the mechanisms, that reduce $q$ and that were in play at each stage of the dynamical evolution of the test asteroids for all six ERs, is given in Table~\ref{tab:percentages}. By comparing the recorded ER and the dynamical mechanism that brought each test asteroid below $0.4\au$ and below $0.076\au$, we find that 400 test asteroids were trapped in the 2:1J MMR throughout their entire evolution since they escaped the asteroid belt until they reach close to the Sun. 3,181 test asteroids spent their entire lifetimes as NEAs trapped in the 3:1J MMR, 47 in the 5:2J MMR, and 3,284 the $\nu_6$ secular resonance. Again, for the last recorded mechanism and the mechanism at $q=q_l$, we split asteroids flagged as being in both the $\nu_5$ and $\nu_6$ resonances, or the $\nu_3\nu_4$ and $\nu_2$ resonances, equally in the respective subgroups.

\begin{table*}
\centering
\caption{An overview of number distribution of resonant mechanisms that have been responsible for bringing the $q$ of asteroids below $q_l$ (second column) and $q^*$ (fourth column) for test particles that entered the near-Earth region through each ER, as well as the predominant mechanism during $\tau_\text{lq}$ (third column). We only include the information for the most important resonances.}
\label{tab:percentages}

\begin{subtable}{0.33\textwidth}
\centering
\caption{2:1J complex}
\begin{tabular}{cccc}
\hline
mechanism & $N_{q_l}$ & $N_{\tau_{lq}}$ &  $N_{q^*}$  \\
\hline
$\nu_5$      & 345  & 343  & 737 \\
2:1J MMR     & 1093 & 1082 & 560 \\
unidentified+Kozai          & 268  & 340  & 353 \\
$\nu_6$      & 51   & 42   & 110 \\
$\nu_3\nu_4$ & 46   & 36   & 54  \\
$\nu_2$      & 27   & 18   & 41  \\
3:1J MMR     & 24   & 30   & 28  \\
5:2J MMR     & 37   & 33   & 12  \\
other        & 58   & 25   & 54 \\ 
\hline
total &  & 1949  &   \\
\end{tabular}
\end{subtable}
\hfil
\begin{subtable}{0.33\textwidth}
\centering
\caption{3:1J complex}
\begin{tabular}{cccc}
\hline
mechanism & $N_{q_l}$ & $N_{\tau_{lq}}$ &  $N_{q^*}$  \\
\hline
3:1J MMR     & 5645  & 5017  & 4054 \\
$\nu_6$      & 1191  & 1493  & 2486 \\
$\nu_5$      & 282   & 526   & 985  \\
unidentified+Kozai          & 1022  & 1265  & 502  \\
$\nu_3\nu_4$ & 538   & 456   & 494  \\
4:1J MMR     & 104   & 64    & 192  \\
$\nu_2$      & 45    & 43    & 127  \\
5:2J MMR     & 67    & 55    & 38   \\
other        & 57    & 32    & 73  \\ 
\hline
total &  & 8951  &   \\
\end{tabular}
\end{subtable}
\hfil
\begin{subtable}{0.33\textwidth}
\centering
\caption{5:2J complex}
\begin{tabular}{cccc}
\hline
mechanism & $N_{q_l}$ & $N_{\tau_{lq}}$ &  $N_{q^*}$  \\
\hline
$\nu_5$      & 432  & 599 & 851  \\
$\nu_6$      & 496  & 536 & 747  \\
3:1J MMR     & 718  & 716 & 635  \\
unidentified+Kozai          & 557  & 621 & 286  \\
$\nu_3\nu_4$ & 253  & 170 & 160  \\
5:2J MMR     & 339  & 218 & 102  \\
$\nu_2$      & 20   & 22  & 42   \\
4:1J MMR     & 18   & 11  & 45   \\
2:1J MMR     & 29   & 31  & 37   \\
7:3J MMR     & 35   & 19  & 18   \\
8:3J MMR     & 46   & 3   & 14   \\
other        & 9   & 6  & 15   \\ 
\hline
total &  & 2952  &   \\
\end{tabular}
\end{subtable}

\vspace{\baselineskip}

\begin{subtable}{0.33\textwidth}
\centering
\caption{$\nu_6$ complex}
\begin{tabular}{cccc}
\hline
mechanism & $N_{q_l}$ & $N_{\tau_{lq}}$ &  $N_{q^*}$  \\
\hline
$\nu_6$      & 4374  & 5606  & 6515 \\
$\nu_3\nu_4$ & 2976  & 2711  & 3034 \\
4:1J MMR     & 2559  & 1834  & 2103 \\
3:1J MMR     & 1915  & 1883  & 1769 \\
$\nu_5$      & 644   & 867   & 1326 \\
unidentified+Kozai          & 3663  & 3424  & 1222 \\
$\nu_2$      & 94    & 41    & 310  \\
5:1J MMR     & 66    & 12    & 90   \\
7:2J MMR     & 155   & 96    & 78   \\
other        & 36    & 8     & 35   \\ 
\hline
total &  & 16482  &   \\
\end{tabular}
\end{subtable}
\hfil
\begin{subtable}{0.33\textwidth}
\centering
\caption{Hungarias}
\begin{tabular}{cccc}
\hline
mechanism & $N_{q_l}$ & $N_{\tau_{lq}}$ &  $N_{q^*}$  \\
\hline
$\nu_6$      & 954   & 1255 & 1504 \\
$\nu_5$      & 1151  & 1426 & 1482 \\
4:1J MMR     & 1581  & 960  & 1356 \\
$\nu_3\nu_4$ & 834   & 728  & 810  \\
3:1J MMR     & 348   & 374  & 595  \\
unidentified+Kozai          & 1545  & 1785 & 544  \\
$\nu_2$      & 88    & 78   & 178  \\
5:1J MMR     & 73    & 13   & 113  \\
7:2J MMR     & 43    & 16   & 21   \\
other        & 27    & 9    & 41   \\ 
\hline
total &  & 6644  &   \\
\end{tabular}
\end{subtable}
\hfil
\begin{subtable}{0.33\textwidth}
\centering
\caption{Phocaeas}
\begin{tabular}{cccc}
\hline
mechanism & $N_{q_l}$ & $N_{\tau_{lq}}$ &  $N_{q^*}$  \\
\hline
3:1J MMR     & 3606  & 3293 & 3647 \\
$\nu_5$      & 635   & 1229 & 1750 \\
4:1J MMR     & 1187  & 554  & 1053 \\
$\nu_6$      & 198   & 290  & 382  \\
unidentified+Kozai          & 1198  & 1763 & 323  \\
$\nu_3\nu_4$ & 627   & 402  & 253  \\
$\nu_2$      & 146   & 138  & 179  \\
5:2J MMR     & 68    & 47   & 56   \\
5:1J MMR     & 17    &  4   & 42   \\
7:2J MMR     & 39    &  8   & 20   \\
other        & 32    & 25   & 48   \\ 
\hline
total &  & 7753  &   \\
\end{tabular}
\end{subtable}

\end{table*}

\begin{figure*}
\centering
\includegraphics[width=\textwidth]{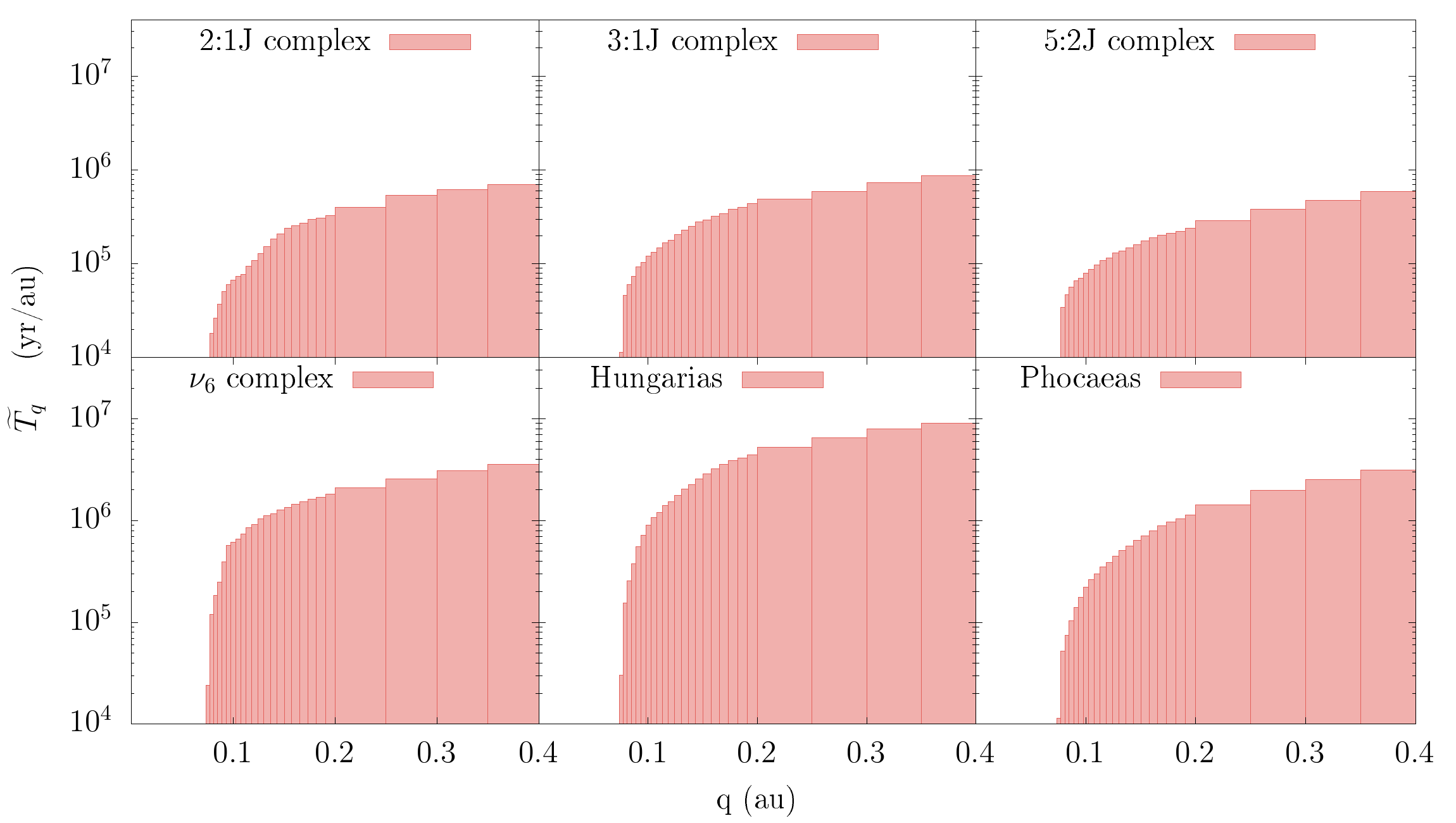}

\includegraphics[width=\textwidth]{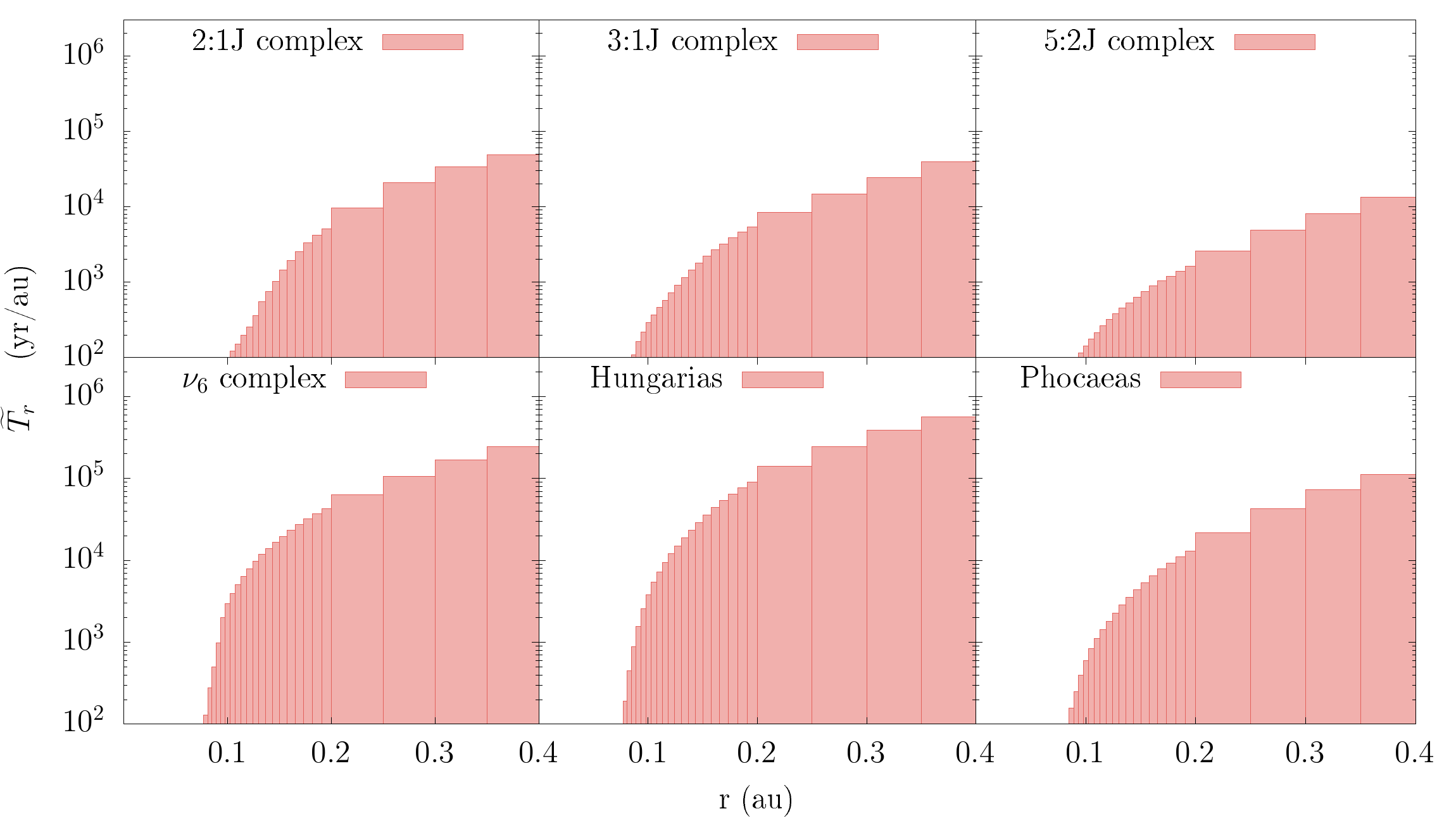}

\caption{ The time asteroids spend at different $q$ intervals (top panels) and at different heliocentric distance $r$ (bottom panels) averaged over test asteroids originating from each ER and normalized by bin width. }
\label{fig:hist_bins_ER_dis}
\end{figure*}

\begin{figure*}
\centering

\includegraphics[width=\textwidth]{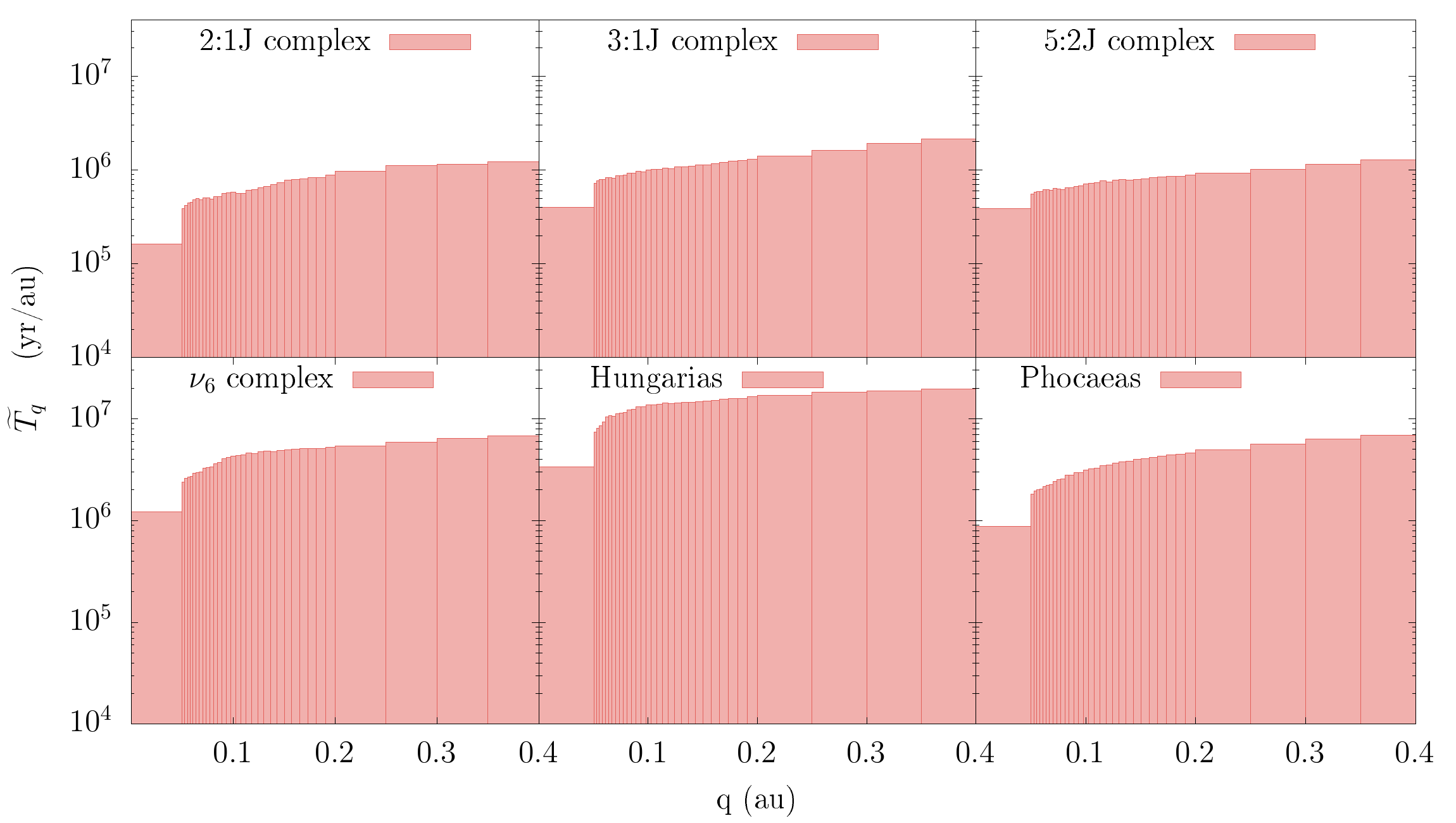}

\includegraphics[width=\textwidth]{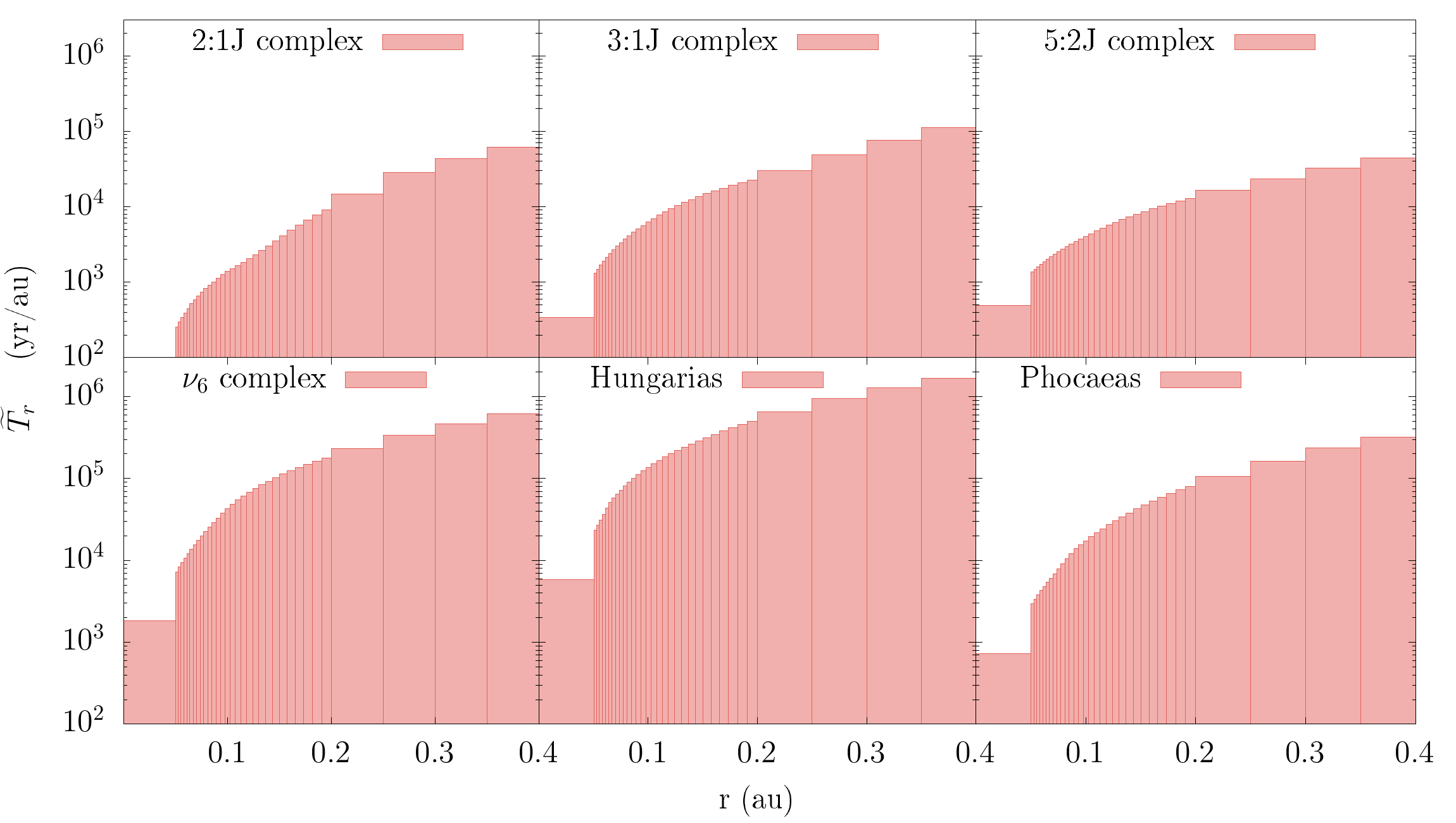}

\caption{The same as in Fig.~\ref{fig:hist_bins_ER_dis} but without taking into account a total disintegration of asteroids at $q^*$.}
\label{fig:hist_bins_ER}
\end{figure*}

We calculated the average dwell time of asteroids with different $q$ and $r$, similarly to the work done in Sec.~\ref{sec:rel_qr}, following the method described in Sec.~\ref{sec:methods} (Fig.~\ref{fig:hist_bins_ER_dis}). We then repeated the same process but without considering a total disruption of test asteroids below $q^*$. In Fig.~\ref{fig:hist_bins_ER}, we show the resulting histograms following the same classification of test asteroids according to ER. $\widetilde{T_q}$ and $\widetilde{T_r}$ are evidently larger in this case, as we do not stop taking into account any subsequent evolution after $q$ passes below the critical value. 

We note that the fact that our model assumes an instantaneous disintegration at $q^*$ may lead to an underestimation of the time that real asteroids spend at different $q$ and $r$. It could be that the mechanism that destroys asteroids close to the Sun requires a relatively long time before an asteroid totally disintegrates \citep{2021Icar..36614535M}. Consequently, asteroids with large variations in $q$ and that therefore spend relatively small amounts of time at small $q$ might, in our model, appear to meet their demise earlier than they do in reality. 

\section{Discussion}

\subsection{Relative importances, time-scales, and efficiencies of resonances that reduce $q$}

Identifying the relevant mechanisms that drive the dynamical evolution of NEAs is not a trivial task. During their dynamical lifetimes, asteroids may enter or leave mean motion and secular resonances, significantly affecting their $e$. In addition, the Kozai mechanism results in large variations in the $e$, potentially driving their $q$ close to the Sun. 

Qualitatively, we identified the same evolutionary paths to near-Sun orbits as previous studies \citep{Farinella1994,Froeschle1995,Gladman1997,Gladman2000,Foschini2000}. While the previous studies focused on MMRs with Jupiter and Saturn, we also considered MMRs with Earth. However, we found that these latter MMRs are not important in decreasing the $q$ of NEAs below $q^*$. 

Contrary to the previous studies, we are able to make quantitative estimates of the relative importance of different resonant mechanisms as well as their time-scales and efficiencies. We found that the majority of the test asteroids that we considered crossed the $q=q_l=0.4\au$ threshold trapped in the 3:1J MMR. In the final stage, when $q^*<q<q_l$, 3:1J MMR and $\nu_6$ were the most important resonances. We also found test asteroids to be trapped in other resonances, such as $\nu_5$, $\nu_3\nu_4$, and the 4:1J MMR, during their evolution in the range $q^*<q<q_l$. A non-negligible portion of the test asteroids are not trapped in a resonance during their small-$q$ evolution until the very last stages which then results in their $q$ becoming smaller than $q^*$. The resonances that give the final `push' below that threshold are mostly $\nu_6$ and the 3:1J MMR, followed by $\nu_5$, $\nu_3\nu_4$ and the 4:1J MMR. It is interesting that for $\sim7$ per cent of our test asteroids, no resonance has been identified to drive $q<q^*$, meaning that the oscillations in $e$ due to the Kozai mechanism or some other dynamical mechanism was sufficient to do so (Table \ref{tab:mechanisms}).

One of the primary goals of this study is to compute $\tau_\text{lq}$, the effective dynamical small-$q$ lifetimes of NEAs ($q^*<q<q_l$), classified according to the resonances they are trapped in. \citet{Gladman1997,Gladman2000,Foschini2000} suggest that the typical dynamical lifetimes of NEAs are $\sim10^7\yr$. In particular, \citet{Gladman1997} calculated the time-scales of the half-life decay of active particles according to each resonance, and found it to be between 2-2.5 Myr for $\nu_6$ and 3:1J MMR, $\sim0.5$ Myr for 5:2J and much longer for the rest 8:3J, 7:3J, 9:4J and 2:1J MMRs. \citet{Farinella1994} found that, in general, near-Sun NEAs trapped in resonances have lifetime of the order of $10^6\yr$, while 3:1J MMR may drastically raise the eccentricity of an NEA in $<10^5\yr$. \citet{Jopek1995} suggest that the objects they studied collided with the Sun within a few $10^5\yr$, while \citet{Foschini2000} argue that the recorded dynamical life times range from $10^5\yr$ to as long as $>10^7\yr$.

All these studies give an estimate on the expected time that an asteroid trapped in a resonance can be found having an orbit in the near-Earth region. However, we are more interested in the time-span during which an NEA can have an orbit that reaches below the orbit of Mercury, and how this differs according to each resonance. We found that MMRs are the faster acting resonances, quickly reducing the $q$ below $q^*$. The median of $\tau_\text{lq}$ for the 4:1J MMR is $\sim7.6 \times 10^4\yr$ , for 3:1J MMR $\sim9.5 \times 10^4\yr$, for $\nu_6$ $\sim2.1 \times 10^5\yr$, for for $\nu_3\nu_4$ $\sim5.6 \times 10^5\yr$, while for $\nu_5$ it is the longest: $\sim2.3 \times 10^6\yr$. In addition, after measuring the rate of change of $q$ of test asteroids during their small-$q$ evolution, we found that the MMRs 3:1J and 4:1J are the most efficient in reducing the $q$ of our test asteroids, according to the metric that we defined.

Furthermore, we have been able to compute the time asteroids spend, on average, at different $q$ and $r$ from the Sun. This is a very relevant question that must be answered if one aims to estimate the disruption rates of asteroids, caused by the possibly thermal mechanism under investigation \citep{2022P&SS..21705490T,2021PSJ.....2..165M,2022Icar..38114995L}. As an example, in the case of the 3:1J MMR, focusing on the range $[q^*,0.2\au]$, we can divide $\widetilde{T_q}$ for each bin with the width of the bin, which gives the dwell time of $q$ in some small range $\Delta q$, comparable to the bin width. We can then make a linear approximation and get a model of the evolution of the dwell time as a function of $\Delta q$ in this range, $t_\text{dwell}(q)=14558.39\Delta q-1042.16\,$yr.     

\subsection{Distribution of albedos among different mechanisms that reduce $q$}

In Sec.~\ref{sec:ERS}, we presented the distribution of mechanisms that reduce $q$ and that were predominant during the small-$q$ evolution of test asteroids, originating from each of the six ERs defined in the Granvik NEO population model. By combining this information with the estimated contribution of each ER and their corresponding albedo probabilities $p_\text{ER}$ \citep{Morby2020}, we can estimate the ratio of dark vs. bright asteroids that were delivered close to the Sun by the resonances considered. Although \citet{Morby2020} suggest three categories of asteroids depending on their albedos ($p_V$), in this study, we consider dark asteroids that have $p_v \leq 0.1$ and bright those with $p_V > 0.1$.

\citet{Granvik2018} provide the values of the relative fraction of NEAs coming from each ER, $\beta_\text{ER}$. Thus, for every resonance, we calculate the weighted contribution of each ER. Note that we have not considered Jupiter family comets (JFCs), so the contributions of the six ERs have been normalized for the absence of JFCs.

\citet{Morby2020} provide the fraction of asteroids with low and high albedos, $p_\text{ER}$, originating from each of the ERs. By multiplying the normalised weighted contribution from each ER with the $p_\text{ER}$, we get the probability that a certain mechanism drives dark or bright asteroids very close to the Sun, according to its contribution to the whole small-$q$ NEA population.

\begin{figure}
\centering
\includegraphics[width=0.48\textwidth]{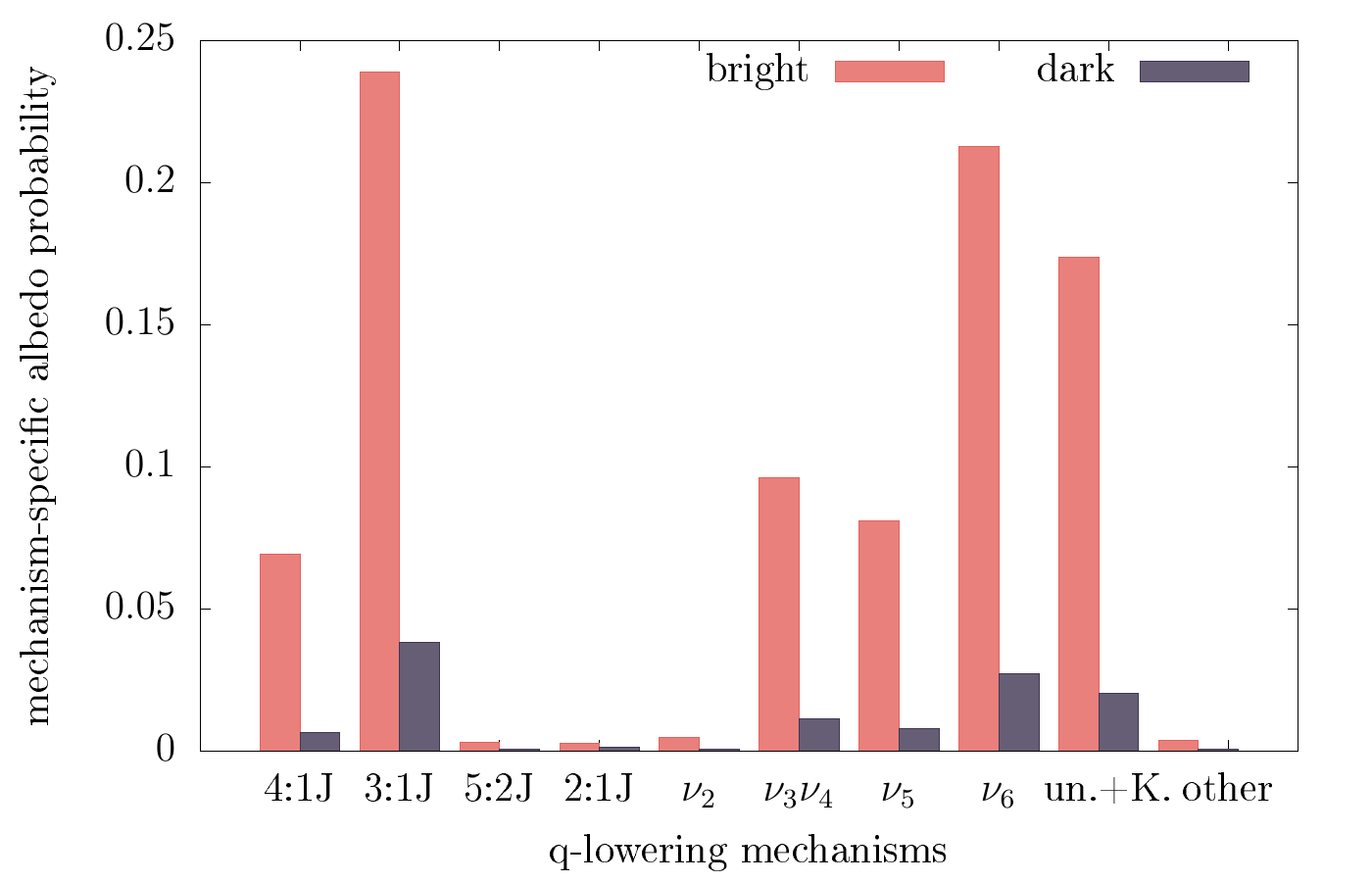}
\caption{ The mechanism-specific albedo probabilities $p_\text{ER}$, provided by \citet{Morby2020} weighted according to \citet{Granvik2018}, that give the percentage of bright and dark asteroids that each predominant mechanism during the small-$q$ evolution of test asteroids contributes to the NEA population. Note that only asteroids that eventually reached below the average disruption distance $q^*$ were considered.}
\label{fig:albedos}
\end{figure}

In Fig.~\ref{fig:albedos}, we show the percentages of bright and dark asteroids that each mechanism contributes to the population of NEAs that eventually reach below the average disruption distance $q^*$, excluding JFCs. It is apparent that very few of these test asteroids have low albedos, corresponding to primitive asteroids of C, P and D spectral classes. However, considering JFCs, which contribute $\sim 8$ per cent of the total near-Earth object (NEO) population, this number might become a little larger. This, however, means that there is a dynamical preference for bright asteroids reaching small-$q$ orbits. In fact, we find that 89 per cent of the test asteroids that reach $q<q^*$ are bright. Note that, since in this analysis we only consider a sample of test asteroids---those that eventually get very close to the Sun ($q<q^*$)---we have included a bias. Consequently, we cannot make predictions for any real asteroid with $q^*<q<q_l$ found trapped in a resonance, because that asteroid may never reach $q<q^*$.

Next, we investigate the possibility that there is a correlation between the small-$q$ lifetime $\tau_\text{lq}$ and the albedo of test asteroids. To that end, we consider the albedo probability distribution functions (PDFs) of asteroids to be described by two distinct Rayleigh distributions:
\begin{equation}\label{eq:PDF}
P(x)=\frac{x e^{\frac{-x^2}{2\sigma^2} }}{\sigma^2}\,,
\end{equation}
one for dark and one for bright asteroids respectively. The choice of $\sigma$ for each albedo group is adopted from \citet{Wright2016}, as it provides an appropriate PDF shape; $\sigma_d=0.03$ and $\sigma_b=0.168$ for dark and bright asteroids respectively. Each test asteroid in our dataset, denoted as coming from a specific ER, must be assigned a random albedo. This is done by taking the inverse of the cumulative distribution function (CDF) of Eq.~\ref{eq:PDF}.
\begin{equation}\label{eq:albedo}
p_V=\sigma\sqrt{-2 ln(U)}\,,
\end{equation}
where $U$ is a uniform random number distribution between 0 and 1 and $\sigma$ is equal to $\sigma_d$ for dark, or $\sigma_b$ for bright asteroids. The number of dark asteroids versus the number of bright asteroids for each ER is decided to be in accordance to the $p_\text{ER}$ values from \citet{Morby2020}. For example, for the 3:1J ER, we selected 1289 albedo values in the range $0<p_V\le 0.1$ from the Rayleigh PDF with $\sigma=\sigma_d$ and 7662 albedo values in the range $0.1<p_V<1$ from the Rayleigh PDF with $\sigma=\sigma_b$. Next, we randomly assign these values to the 8951 asteroids coming from the 3:1J ER, that eventually reached $q<q*$. We repeat the same process for all ERs.

Next, we search for a potential correlation between $p_V$ and $\tau_\text{lq}$. We calculate Pearson's correlation coefficient weighted according to $\beta_\text{ER}$:
\begin{equation}\label{eq:Pearson's}
\text{corr}(p_V,\tau_\text{lq},\beta_\text{ER})=\frac{\text{cov}(p_V,\tau_\text{lq},\beta_\text{ER})}{\sqrt{\text{cov}(p_V,p_V,\beta_\text{ER})\text{cov}(\tau_\text{lq},\tau_\text{lq},\beta_\text{ER})}}\,,
\end{equation}
where the covariance is given by
\begin{equation}\label{eq:covariance}
\begin{split}
\text{cov}(p_V,\tau_\text{lq},&\beta_\text{ER})=  \\ & \frac{\sum_{i=1}^{N_\text{ast}} \beta_{\text{ER}_i} (p_{V_i}-m(p_V,\beta_\text{ER}))(\tau_{\text{lq}_i}-m(\tau_\text{lq},\beta_\text{ER})) } {\sum_{i=1}^{N_\text{ast}} \beta_{\text{ER}_i} } \,,
\end{split}
\end{equation}
and respectively, for $\text{cov}(p_V,p_V,\beta_\text{ER})$ and $\text{cov}(\tau_\text{lq},\tau_\text{lq},\beta_\text{ER})$. The weighted mean is calculated by
\begin{equation}\label{eq:w_mean}
m(p_V,\beta_\text{ER}))=\frac{\sum_{i=1}^{N_\text{ast}} \beta_{\text{ER}_i} p_{V_i}}{\sum_{i=1}^{N_\text{ast}} \beta_{\text{ER}_i} \tau_{\text{lq}_i}}\,,
\end{equation}
and accordingly we can also get $m(\tau_\text{lq})$. 

Taking into account all test asteroids originating from each one of the six ERs, we find a correlation coefficient of 0.01, which implies that there is hardly any correlation between albedo and low-$q$ lifetime of asteroids.  This result suggests that the fraction of 89 per cent bright and 11 percent dark asteroids in the near-Earth region is not affected by the resonant mechanisms dominating their dynamical evolution. 

\citet{Morby2020} suggest that the fraction of NEOs with $p_V\le0.1$ is 30 per cent for asteroids with $q<0.2\au$, which is the size of the $q$ bin in their analysis. The apparent discrepancy (11 versus 30 per cent) can be explained considering that our sample of test asteroids only contains those that eventually reach $q<q^*$, reducing the number of dark asteroids mainly originating from the outer asteroid belt. E.g., a low-inclination test asteroid with $a\gtrsim2.6\au$ cannot acquire $q<q^*$ before it typically suffers a close encounter with Jupiter and is ejected from the inner Solar System. Hence, most of the test asteroids that reach $q<q^*$ will have $a\lesssim2.6\au$. For $q<0.2\au$ the range in semimajor axis of those objects that are ejected due to encounters with Jupiter increases to $a\gtrsim2.7\au$. As seen in Figs.~2 and 7 of \citet{Morby2020}, the albedo distribution becomes darker when going from $a\sim2.6\au$ to $a\sim2.7\au$. This explains, at least partly, the apparent discrepancy between our results and those presented by \citet{Morby2020}.

\subsection{Implications for the EKL mechanism}

In the test particle limit, as discussed in Sec.~\ref{sec:lib_mech}, the EKL mechanism can lead to 'flips' of test asteroids such that their orbital direction changes from prograde ($i<90\deg$) to retrograde ($i>90\deg$), or the other way round. During such a flip, the $e$ of the test asteroid increases to an extremely large value. That is because the strength of the coupled $e$ and $i$ oscillations are very sensitive to the value of the test asteroid's $i$, and, as it approaches $90\deg$, arbitrarily strong oscillations occur. However, in this study, we witness very few flips. In fact, we identified 91 test asteroids which had their orbits flip from prograde to retrograde, for a total of 281 flips. For smaller $a$, as $e$ increases, the $q$ of an asteroid becomes smaller than $q^*$. Since we disregard any future evolution, we fail to see the change in the orbital plane. For larger $a$, as $e$ increases, the aphelion of an asteroid intersects the orbit of Jupiter and can then get ejected from the inner Solar System, removing it from our sample of test asteroids. In addition, \citet{Naoz2013b} suggest that general relativity, which was ignored in the orbital integrations for the \citet{Granvik2016,Granvik2018} NEO models, suppresses inclination flips if the corresponding post-Newtonian time-scale is much smaller than the Newtonian quadrupole time-scale. As a result, even though the number of flips is already small, the simulations have a positive bias for asteroids with retrograde orbits, and the fraction of real asteroids having undergone flips is likely to be smaller than the simulations suggest. The fact that the only retrograde NEA known to date was discovered 13 years ago, supports the conclusion that flips are rare. In fact, \citet{Greenstreet2012} suggest that NEAs on retrograde orbits are most likely generated by the Kozai mechanism and the 3J:1 MMR, a result later supported also by \citet{Granvik2018}.

\begin{figure}
\centering
\includegraphics[width=0.48\textwidth]{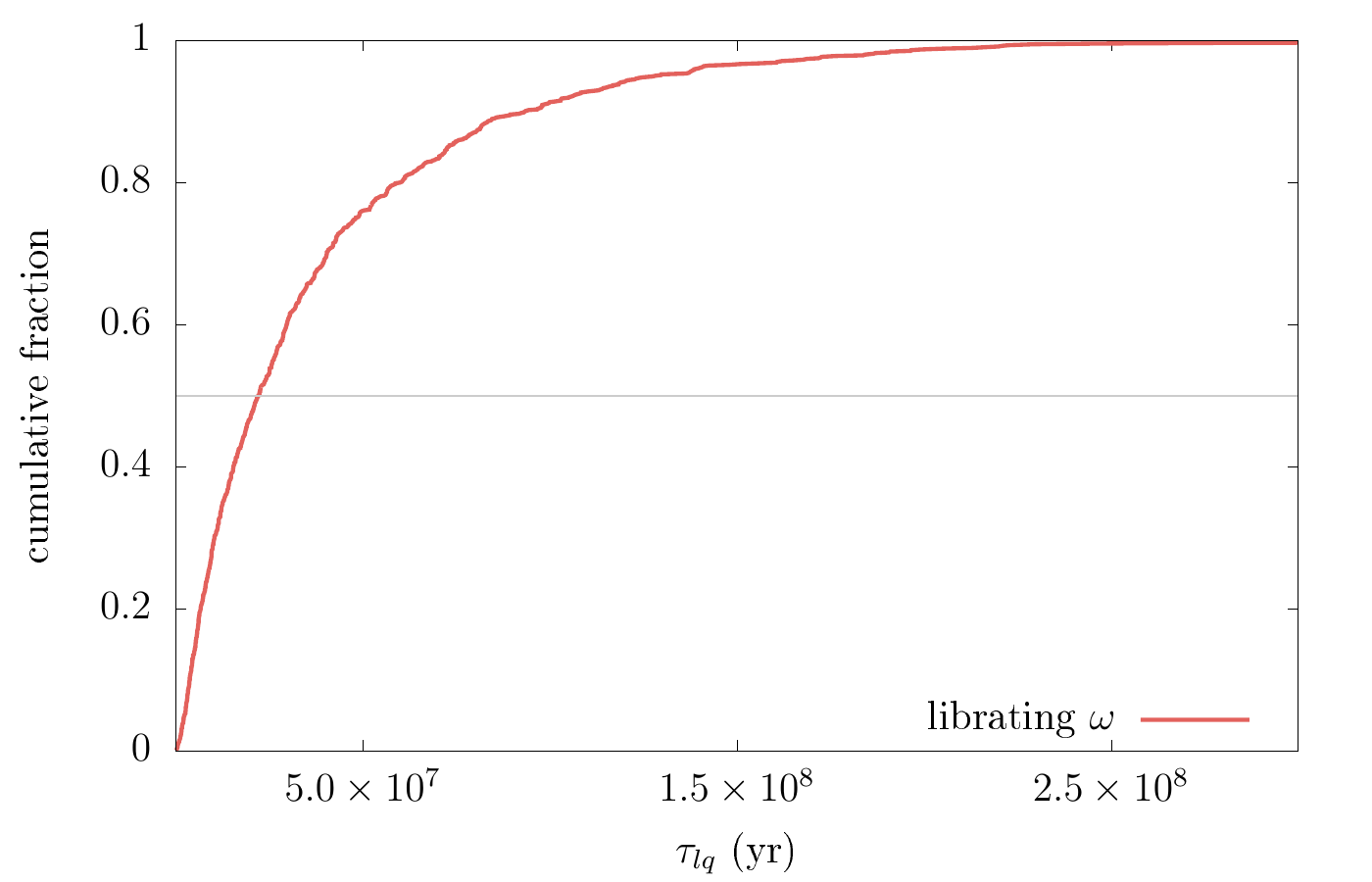}
\caption{The cumulative distribution of the small-$q$ time-scale for test asteroids for which $\omega$ was librating around $90\deg$ or $270\deg$ during their $\tau_{lq}$ evolution, when no other resonance was present. The grey line is the median.}
\label{fig:kozai-lib}
\end{figure}

Finally, we note that when an asteroid's $\omega$ librates around $90\deg$ or $270\deg$, it provides protection from node crossings and, as a result, close encounters with the planets. Consequently, NEAs with librating $\omega$ are particularly long lived (Fig.~\ref{fig:kozai-lib}).

\section{Conclusions}

In this study, we have investigated the dynamical mechanisms that are able to pump the $e$ of NEA to values that allow their $q$ to pass below the average disruption distance $q^*=0.076\au$, as defined by \citet{Granvik2016}, by developing an automated way to identify the occurrence of resonances during all stages of their orbital evolution in the near-Earth region. Previous studies have already shed light on the possible evolutionary paths that NEAs follow during their lifetime. We have extended these efforts by utilizing a much larger sample consisting of tens of thousands of test asteroids.

The synthetic NEA population studied exhibits a great variety of orbital evolutions. The main dynamical mechanisms affecting their orbits are mean motion and secular resonances, which are very often overlapping each other. The most important mechanisms are 3:1J and 4:1J MMRs and the $\nu_3\nu_4$, $\nu_5$ and $\nu_6$ secular resonances. Among these mechanisms, the one acting on the fastest time-scale is the 4:1J MMR, while $\nu_5$ is the slowest. However, in the case that an asteroid is driven close to the Sun by a resonance but then escapes, oscillations coming from the Kozai mechanism, classical or eccentric, can be  enough to push $q$ below $q^*$. 

We have also determined that the time NEAs spend at different heliocentric distances range from a few years very close to the Sun to a few hundreds of thousands of years. These results can be combined with thermal experiments studying the rate at which different asteroid materials get destroyed from the solar irradiation at these heliocentric distances. In this way, one can make predictions about the orbital distribution of the real near-Sun NEA population, shed light on the expected physical lifetimes of small-$q$ asteroids, and test different hypotheses trying to explain the super-catastrophic disruption at small $q$. 

Asteroids following escape routes from the inner main belt are the most likely to achieve orbits with small $q$ and are also most probably bright. Even though these bright asteroids can evolve to small-$q$ orbits on very short time-scales due to resonant mechanisms such as the 3:1J MMR, from a purely dynamical perspective, we expect to find less dark asteroids close to the Sun.

In our analysis, we have ignored planetary close encounters which greatly affect the evolution of $q$ by significantly changing $a$. We investigate the effects of this important mechanism in a forthcoming study. In addition, we have not considered the $\nu_{16}$ secular resonance. Furthermore, the choice of a total 'instantaneous' disruption at $q^*$ can be an oversimplification which may underestimate the lifetime of NEAs. However, since we do not yet fully understand the mechanism that destroys asteroids when they get close to the Sun, this assumption is the only safe one we can make. A further improvement of this would be to test various disruption distances that correspond to asteroids with different sizes and repeat the same analysis to test if and how the significance of each resonance is affected.

\section*{Acknowledgements}

We thank the anonymous reviewer for constructive criticism that improved the paper. AT and MG acknowledge funding from the Knut and Alice Wallenberg Foundation, and MG also from the Academy of Finland and the Waldemar von Frenckell Foundation.

\section*{Data Availability}
No new data were generated in support of this research.



\bibliographystyle{mnras}
\bibliography{bibliography} 








\bsp	
\label{lastpage}
\end{document}